\documentclass[preprint2]{aastex}

\usepackage{graphics}
\usepackage{mathptmx}
\usepackage{natbib}
\usepackage[usenames,dvips]{color}

\newcommand{\msun}{$M_\odot$}

\newcommand{\hi}{H\,{\sc i}\rm}

\newcommand{\sii}{Si\,{\sc i}}
\newcommand{\siii}{Si\,{\sc ii}}
\newcommand{\siiii}{Si\,{\sc iii}}

\newcommand{\fei}{Fe\,{\sc i}}

\newcommand{\tii}{Ti\,{\sc i}}

\newcommand{\Teff}{T_{\rm eff}}
\newcommand{\EW}{W_{\lambda}}
\newcommand{\mA}{{\rm m\AA}}
\newcommand{\Elow}{E_{\rm low}}
\newcommand{\Eup}{E_{\rm up}}

 \newcommand{\hii}{\ion{H}{2}}

\slugcomment{To appear in Astrophysical Journal}

\shorttitle{Red Supergiants and J-band NLTE effects}
\shortauthors{Bergemann et al.}

\begin{document}


\title{Red Supergiant Stars as Cosmic Abundance Probes:\\
    II. NLTE Effects in J-band Silicon Lines}


\author{Maria Bergemann}
\affil{Max-Planck-Institute for Astrophysics, Karl-Schwarzschild-Str.1, D-85741
Garching, Germany}
\email{mbergema@mpa-garching.mpg.de}
\author{Rolf-Peter Kudritzki\altaffilmark{1,2}}
\affil{Institute for Astronomy, University of Hawaii, 2680 Woodlawn Drive,
Honolulu, HI 96822}
\email{kud@ifa.hawaii.edu}
\author{Matthias W\"url}
\affil{Max-Planck-Institute for Astrophysics, Karl-Schwarzschild-Str.1, D-85741
Garching, Germany}
\email{Matthias.Wuerl@physik.uni-muenchen.de}
\author{Bertrand Plez}
\affil{Laboratoire Univers et Particules de Montpellier, Universit\'e
Montpellier 2, CNRS, F-34095 Montpellier, France}
\email{bertrand.plez@univ-montp2.fr}
\author{Ben Davies}
\affil{Institute of Astronomy, Univerity of Cambridge, Madingley Road,
Cambridge, CB3 OHA, UK}
\email{bdavies@ast.cam.ac.uk}

\and
\author{Zach Gazak}
\affil{Institute for Astronomy, University of Hawaii, 2680 Woodlawn Drive,
Honolulu, HI 96822}
\email{zgazak@ifa.hawaii.edu}


\altaffiltext{1}{Max-Planck-Institute for Astrophysics,
Karl-Schwarzschild-Str.1, D-85741 Garching, Germany}
\altaffiltext{2}{University Observatory Munich, Scheinerstr. 1, D-81679 Munich,
Germany}


\begin{abstract}

Medium resolution J-band spectroscopy of individual red supergiant stars 
is a promising tool to investigate the chemical composition of the young
stellar population in star forming galaxies. As a 
continuation of recent work on iron and titanium, detailed non-LTE 
calculations are presented to investigate the influence of NLTE on the 
formation of silicon lines in the J-band spectra of red supergiants.
Substantial effects are found resulting in significantly stronger 
absorption lines of neutral silicon in non-LTE. As a consequence, silicon 
abundances determined in non-LTE are significantly smaller than in LTE 
with the non-LTE abundance corrections varying smoothly between -0.4 dex 
to -0.1 dex for effective temperatures between 3400K to 4400K. The 
effects are largest at low metallicity. The physical reasons behind the
non-LTE effects and the consequences for extragalactic J-band abundance
studies are discussed.

\end{abstract}





\section{Introduction}

With their enormous luminosities of $10^{5}$ to $\sim 10^{6}$
L/L${\odot}$ \citep{1979ApJ...232..409H} emitted at infrared wavelengths red
supergiant stars (RSGs) are ideal probes of extragalactic cosmic abundances. The
J-band spectra of RSGs are dominated by strong and isolated atomic lines of
iron, titanium, silicon and magnesium, while the molecular lines of OH,
H$_{2}$O, CN, and CO are weak. In consequence, medium resolution spectroscopy in
this spectral range is sufficient to derive stellar parameters and chemical
abundances of RSGs from these atomic lines. This has been demonstrated recently
by \citet{davies10} (hereinafter DKF10), who introduced a novel technique using
MARCS model atmosphere spectra \citep{gustafsson08} to determine metallicities
with an accuracy of $\sim$ 0.15 dex per individual star. With existing
telescopes and forthcoming new efficient MOS spectrographs such as MOSFIRE at
Keck and KMOS at the VLT the technique can be applied to an investigation of
metallicities of galaxies up to 10 Mpc distance. Even more exciting are the
perspectives of the use of future adaptive optics (AO) MOS IR spectrographs at
the next generation of extremely large telescopes. \citet{evans11} estimate that
with instruments like EAGLE at the E-ELT and IRMS at the TMT it would be
possible to measure abundances of $\alpha$- and iron-group elements of
individual RSGs out to the enormous distance of 70 Mpc.

This is a substantial volume of the local universe containing entire groups 
and clusters of galaxies, for which the formation and evolution could be 
studied through the determination of accurate abundances from individual 
stars. So far, most of our information about the metal content of star 
forming galaxies is obtained from a simplified analysis of the strongest
\hii~region emission lines. As discussed by \citet{kud08,kud12} and
\citet{bresolin09} these ``strong-line methods'' are subject to large systematic
uncertainties, which are poorly understood. In consequence, alternative methods
using stars such as blue supergiants \citep{kud12} or RSGs are highly
desirable. 

We have, therefore, started to investigate the DKF10 J-band method and its 
possible limitations and systematic uncertainties in more detail. Since the 
MARCS model atmosphere spectra are calculated in local thermodynamic 
equilibrium (LTE), an obvious part of such an investigation is the 
assessment of the influence of departures from LTE, which might be 
important because of the extremely low gravities and hence low 
densities encountered in the atmospheres of RSGs. In a first step, we 
have carried out non-LTE (NLTE) line formation calculations in RSG 
atmospheres for iron and titanium and have discussed the consequences
of NLTE-effects for the J-band analysis (\citealt{bergemann12}, 
hereafter Paper I). It was found that NLTE-effects are small for J-band 
iron lines, but significant for titanium. Now we extend this work and 
present detailed NLTE calculations for silicon. There are four strong 
sub-ordinate \sii~lines observed in J-band spectra of RSGs, which provide 
crucial abundance information. We investigate the influence of NLTE 
effects on the formation of these lines.

The paper is structured as follows. In sections 2 we describe
the model atmospheres and the details of the line formation calculations.
Section 3 presents the results: departure coefficients, line profiles and
equivalent widths in LTE and non-LTE and non-LTE abundance corrections for
chemical abundance studies. Section 4 discusses the consequences for the new
J-band diagnostic technique and aspects of future work.

\section{Model Atmospheres and non-LTE Line Formation}

As in Paper I we use MARCS model atmospheres \citep{gustafsson08} for 
the underlying atmospheric structure and the DETAIL NLTE code 
\citep{butler85} for the calculation of NLTE occupation numbers. Line
profiles and NLTE abundance corrections are computed with the
separate code SIU \citep{reetz} using the level departure coefficients from
DETAIL. SIU and DETAIL share the same physics and line lists. For all 
details such as model structure and geometry, background opacities, 
solar abundance mixture, etc. we refer the reader to Paper I.

To assess the importance of NLTE effects for the J-band \sii~lines 
we use a small grid of models computed assuming a stellar mass
of 15 \msun with five effective temperatures (T$_{\rm eff}$ = 3400, 3800,
4000, 4200, 4400K), three gravities ($\log g = 1.0$, $0.0$, $-0.5$ (cgs)), three
metallicities ([Z]$\,\equiv\,$ log Z/Z$_{\odot}$ $= -0.5$, $0.0$, $+0.5$). The 
micro-turbulence is fixed to $\xi_{t} = 2$ km/s. This grid covers the range of
atmospheric parameters expected for RSG's (see DKF10 and Paper I).

\begin{figure*}
\includegraphics[width=1.0\textwidth]{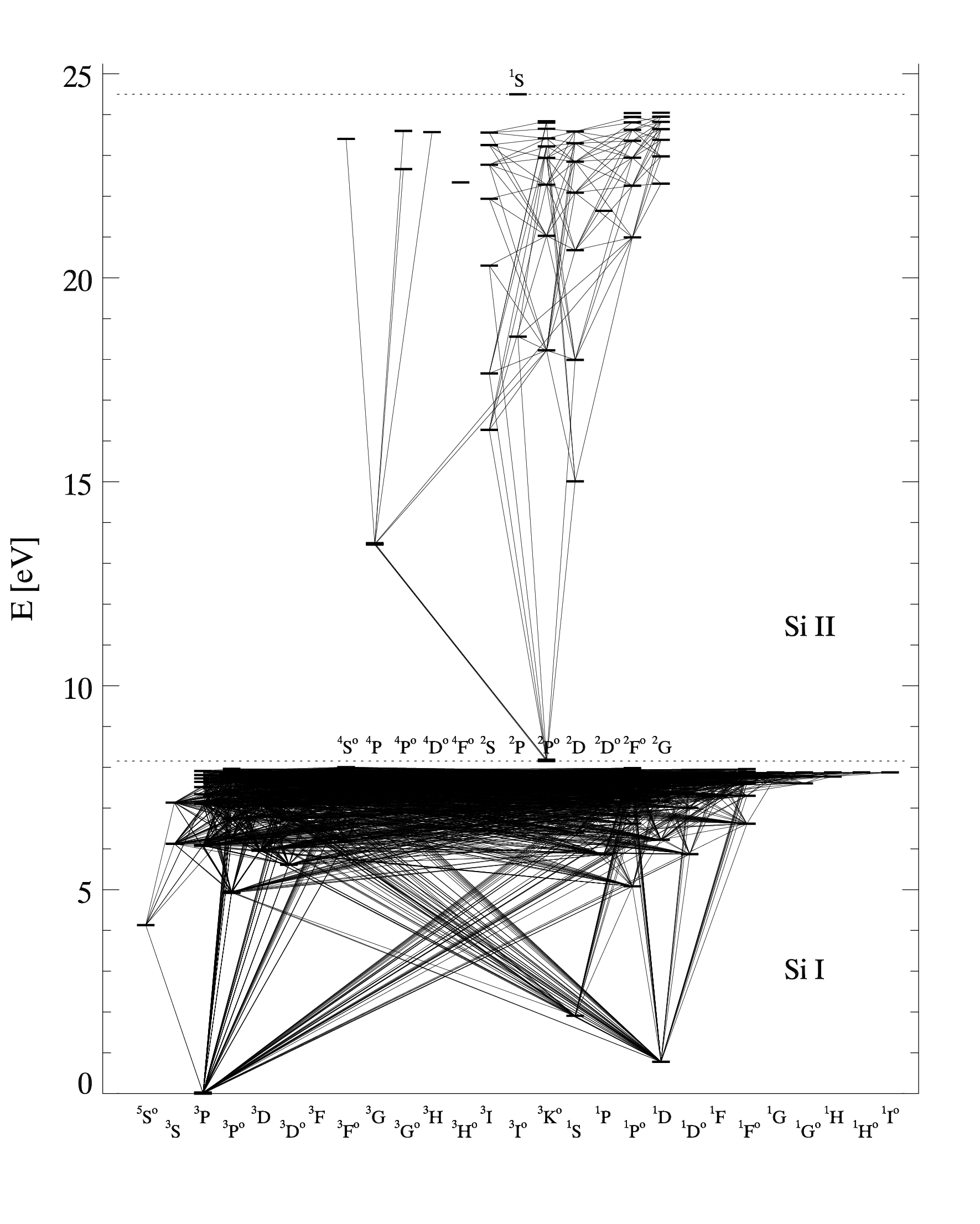}
\caption{The complete silicon atomic model.}
\label{cmod}
\end{figure*}

\begin{figure*}[ht!]
\begin{center}
\includegraphics[scale=0.6,angle=90]{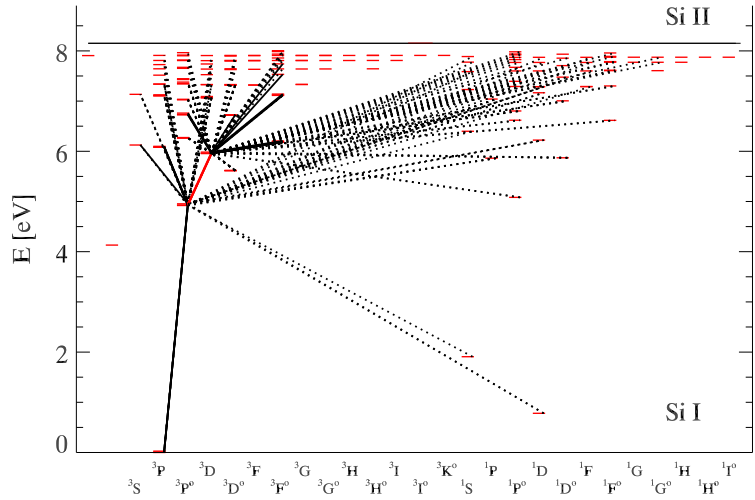}
\caption{The silicon NLTE atomic model showing only transitions to and from
the levels leading to the J-band IR transitions. Strong transition with $\log
gf$-values larger than $-1.0$ are solid, weaker transitions are dotted. The IR
transitions are highlighted in red.}
\label{rmod}
\end{center}
\end{figure*}

\begin{figure*}[ht!]
\begin{center}
\includegraphics[scale=0.5,angle=90]{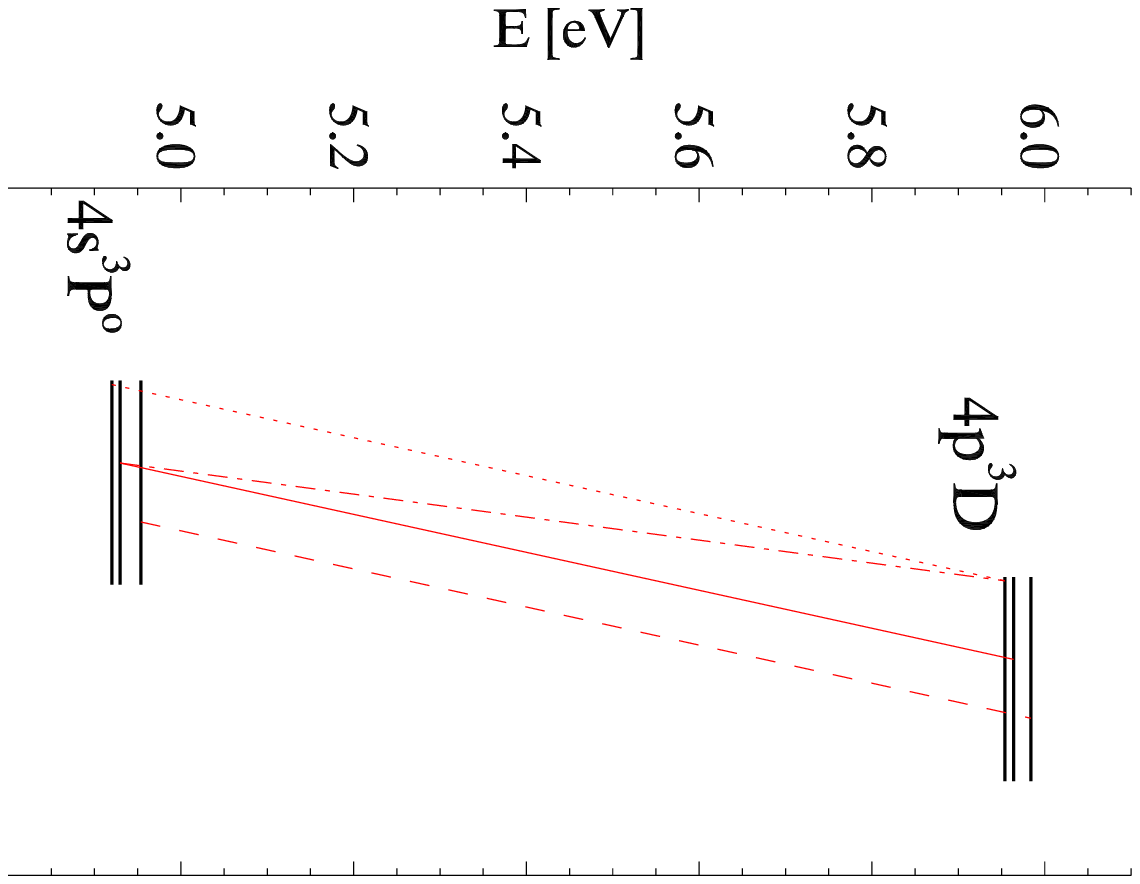}
\caption{Fine structure splitting of the \sii\ IR J-band lines.}
\label{fstruc}
\end{center}
\end{figure*}

\subsection{Model atom}{\label{sec:atom}} 

Our model atom consists of the first three ionization stages of silicon. It has
$289$ levels of \sii\ (mostly singlet and triplet terms and one quintet term), 
$50$ levels of \siii\ (duplet and quartet terms). \siiii\ is represented by 
its ground state only because of the low effective temperatures of RSGs, for
which \sii\ is the dominating ionization stage in the atmospheric layers where
the IR lines investigated in this study form. Fine structure splitting is taken
into account for all \sii\ levels with excitation energies smaller than $7.45$
eV and all \siii\ levels with excitation energies smaller than $6.20$ eV. Most
of the information on energy levels was drawn form the NIST database
\citep{kram12}\footnote{http://www.nist.gov/pml/data/asd.cfm}. Also, $147$
theoretically-predicted \sii\ levels were included from the Kurucz
database\footnote{http://kurucz.harvard.edu/atoms/}. Atomic completeness at
energies close to first ionization threshold is important for statistical
equilibrium of the neutral atom.

The radiative bound-bound transitions between energy levels of \sii\ were 
extracted from the Kurucz database, while the \siii\ data have been taken 
from the NIST database. There is no need for a more refined model of \siii\,
since the ion is un-affected by non-LTE for FGKM stars, which is also confirmed
by our test calculations. In fact, on these grounds the whole \siii\ stage could
be excluded completely. Only transitions between $0.1\mu$m and $30\mu$m
with $\log gf$-values greater than $-8$ were included. In total, the model atom
contains $2956$ allowed transitions, $2826$ for \sii\ and $130$ for \siii\,
respectively. Fig. \ref{cmod} shows the complete atomic model.

While Fig. \ref{cmod} gives an impression about the total effort made in
calculating NLTE effects for the silicon atom, it is not well suited to discuss
the conditions leading to the formation of the IR J-band \sii~lines. For this
purpose, we provide Fig. \ref{rmod}, which shows only transitions from and to
the upper and lower levels of these lines. Fig. \ref{fstruc} zooms into this
figure and displays the corresponding fine structure splitting. The detailed
information about the lines is given in Table 1.

Photoionization cross-sections calculated in the close coupling approximation
using the R-matrix method were taken from the TOPbase database \citep{cunto93} 
and are limited to a minimum of photon wavelengths of 1000\AA. About $20$ \%
of very high levels with excitation energy above $6$ eV were missing in this
database and for those the hydrogenic approximation was adopted.

Electron collisions for excitation and ionization were calculated using the
approximations by \citet{1962ApJ...136..906V} and \citet{1962amp..conf..375S},
respectively, as also given by \citet[][Sections 3.6.1,
3.6.2]{2000asqu.book.....C}. Bound-bound and bound-free collision cross-sections
due to collisions with \hi\ atoms were calculated according to the generalized
formula given by \citet[][A10]{1993PhST...47..186L}. 

\begin{figure}[ht!]
\begin{center}
\resizebox{\columnwidth}{!}{\includegraphics[scale=1,angle=0]
{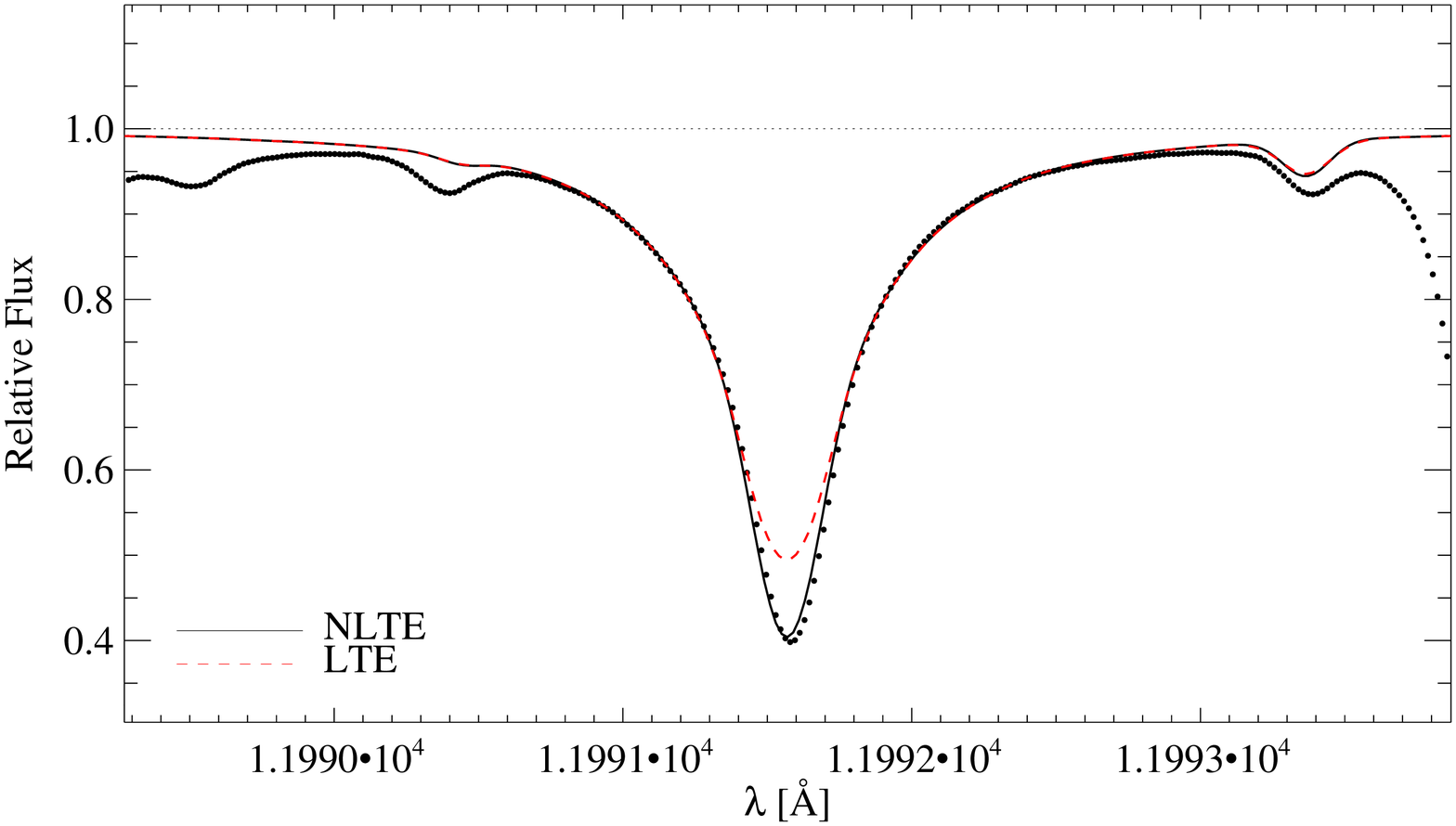}}
\resizebox{\columnwidth}{!}{\includegraphics[scale=1,angle=0]
{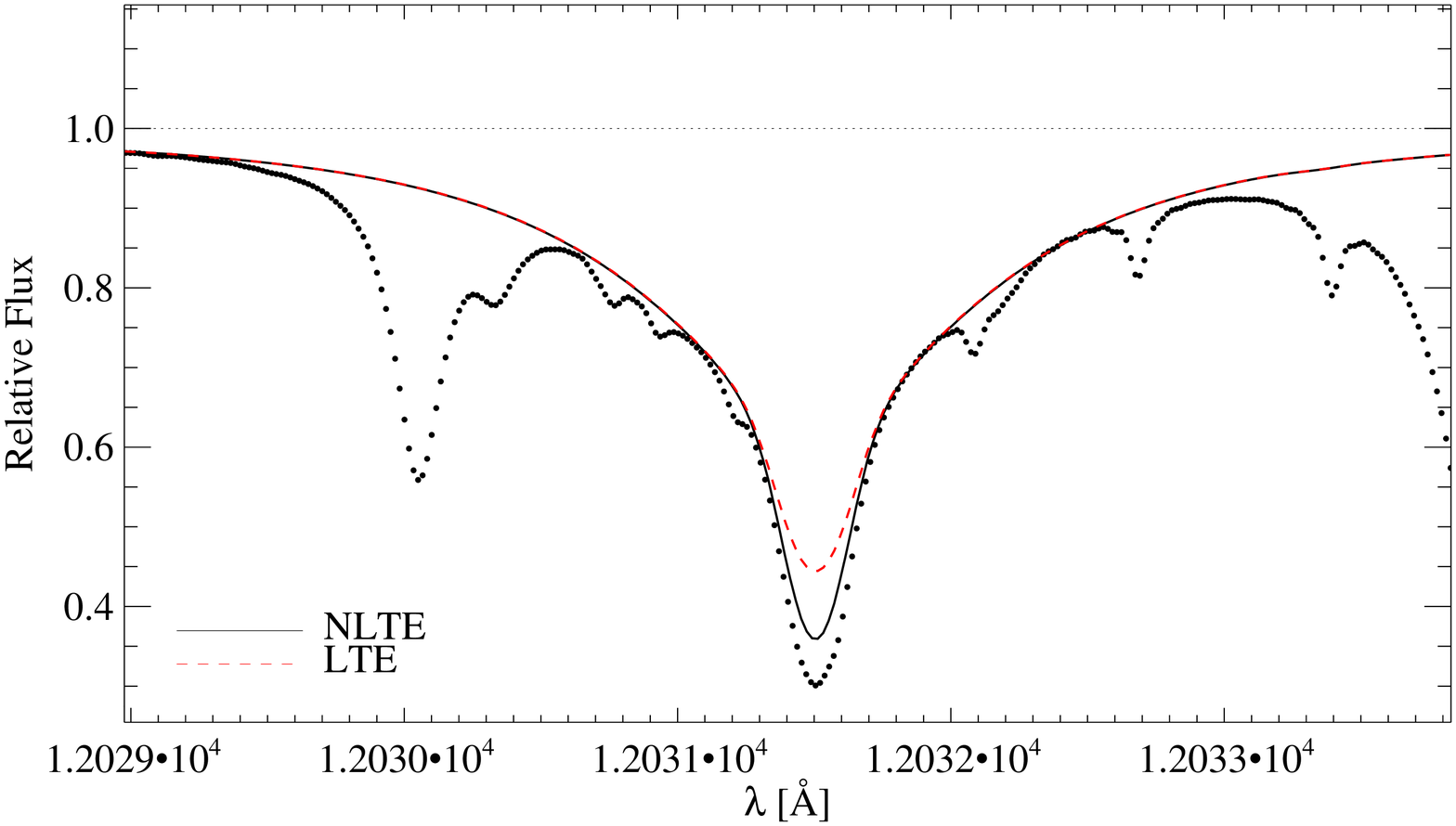}}
\resizebox{\columnwidth}{!}{\includegraphics[scale=1,angle=0]
{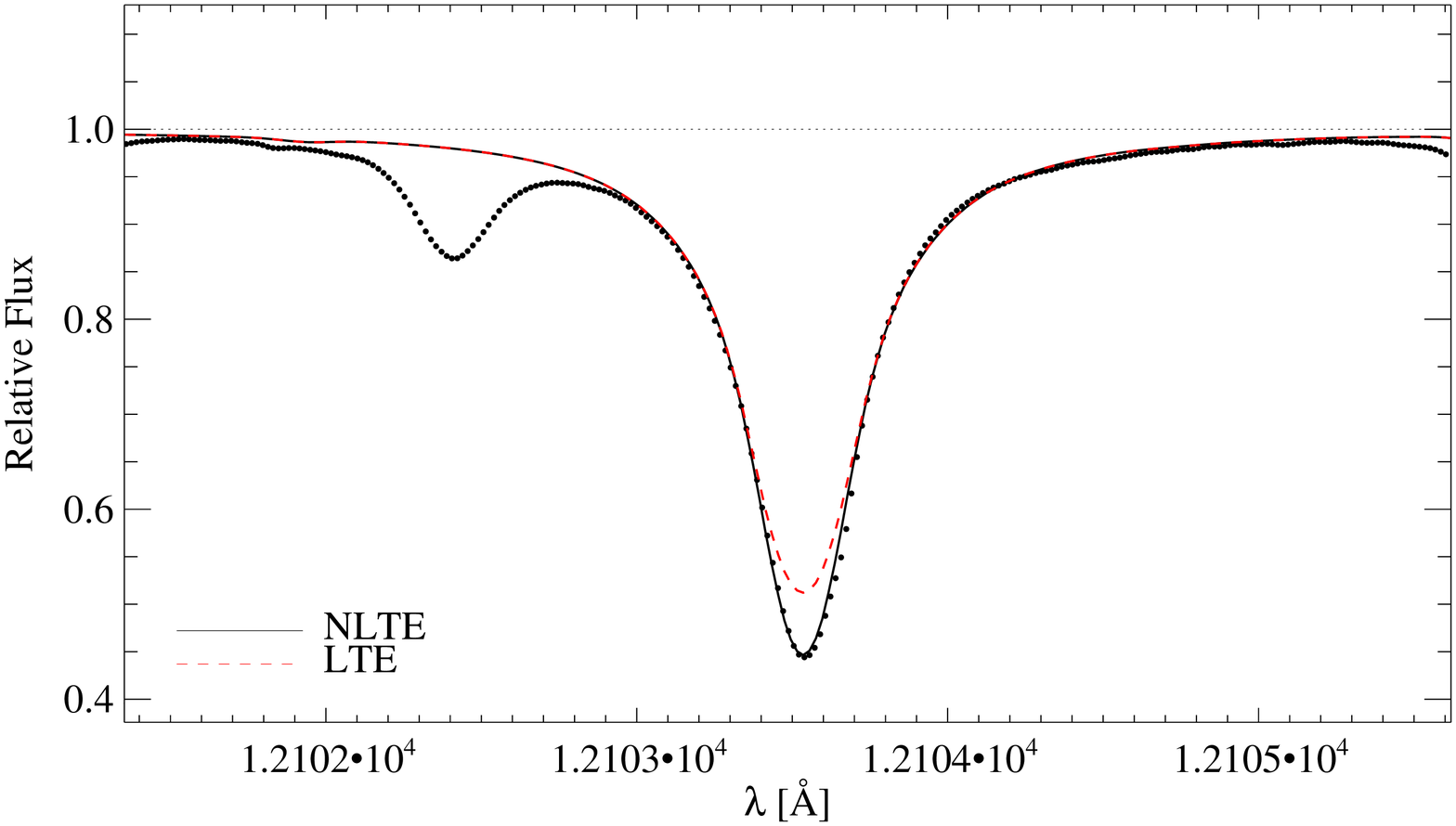}}
\caption{J-band observations of \sii\ lines in the spectrum of the Sun compared
with LTE (red dashed) and non-LTE (black solid) calculations.}
\label{sunir}
\end{center}
\end{figure}

\subsection{Test calculations for the Sun}

As a test of our approach to model the formation of silicon lines we have 
calculated \sii\ and \siii\ lines for the Sun and determined their non-LTE
silicon abundances. For the optical lines, the $gf$ and $C_6$ values were taken
from Shi et al. (2008), and the microturbulence set to $1$ km/s. The result is $\log \rm{A}_{\rm{Si}} = 7.56 \pm{0.05}$ in a very good agreement with the non-LTE analysis by \citet{shi08} and \citet{wedemeyer01}. The NLTE abundance corrections for the lines in common are also fully consistent with the latter investigations: the NLTE effects in the optical \sii\ transitions are minor, typically within $-0.01 \pm 0.01$ dex, but become increasingly important for the IR lines, reaching $-0.1$ dex.

A comparison with the observed silicon IR J-band in the solar KPNO flux spectrum
is given in Fig. \ref{sunir}. The gf-values are that recommended by the VALD2 database, i.e., \citet{kur07}. The non-LTE calculations agree well with the observations in the line cores as well as in the wings. This confirms that the atomic data used for the lines ($gf$-values, radiative and collisional broadening) are reliable. The LTE line profiles are substantially weaker showing that LTE is a poor approximation for the solar IR \sii\ lines and leading to $0.05$ to $0.1$ dex over-estimated abundances. 

Based on the solar analysis, we do not find any need for an arbitrary scaling of the cross-sections in the NLTE calculations, contrary to Shi et al. (2008). They obtain a somewhat better fit for the $12031$ \AA~ \sii\ line adopting a lower efficiency of \hi inelastic collisions. There could be several reasons for that. First, they used a different source of gf-values (Hartree-Fock data from http://nlte.nist.gov/MCHF/). Second, they could have assumed a different value of micro- and macroturbulence. Unfortunately, Shi et al. (2008) do not provide $\xi_{t}$ and $V_{\rm mac}$ in the paper for us to make a detailed comparison.

\begin{figure*}[ht!]
\begin{center}
\resizebox{0.6\textwidth}{!}{\includegraphics[scale=0.8]{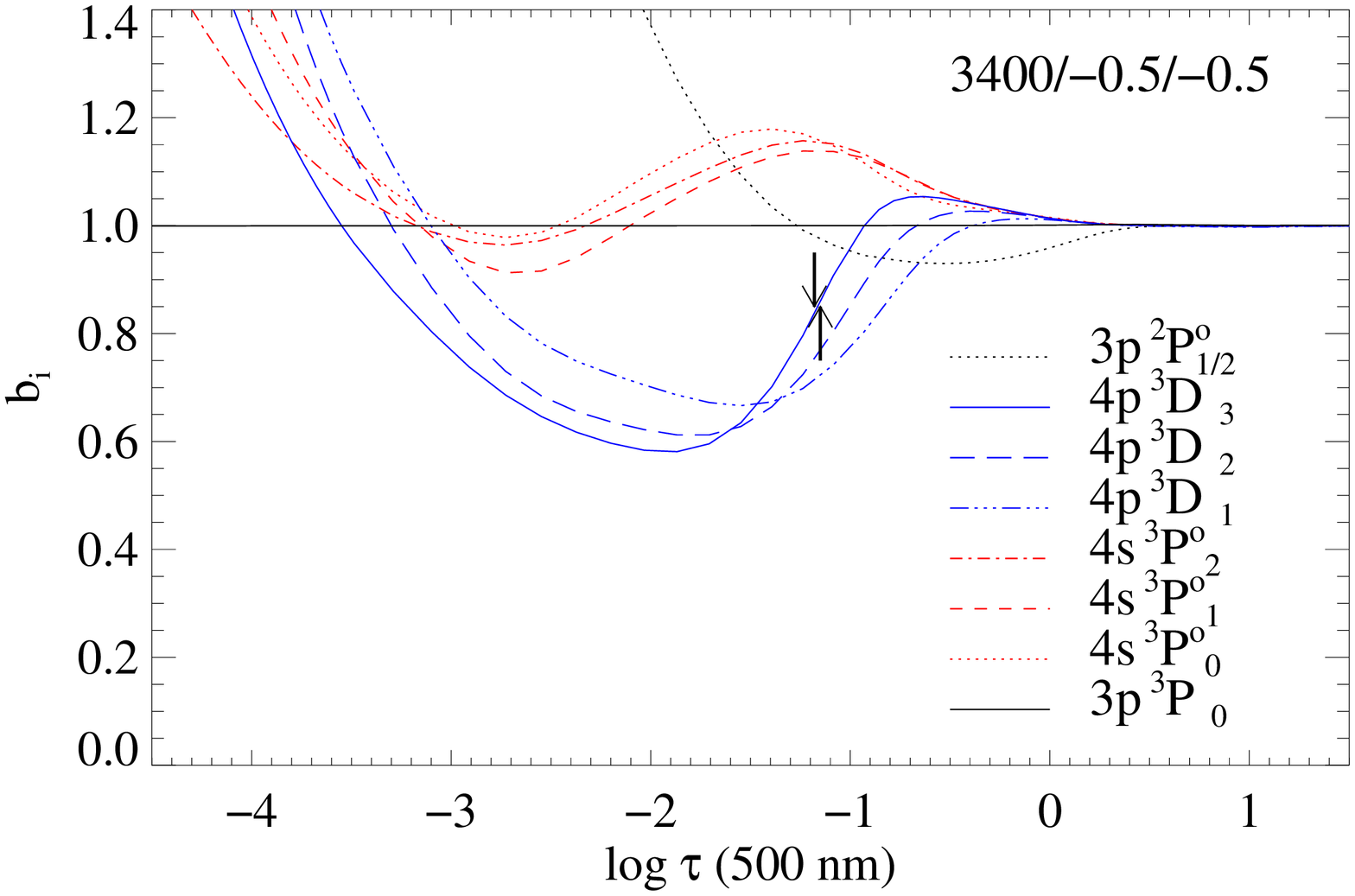}}
\resizebox{0.6\textwidth}{!}{\includegraphics[scale=0.8]{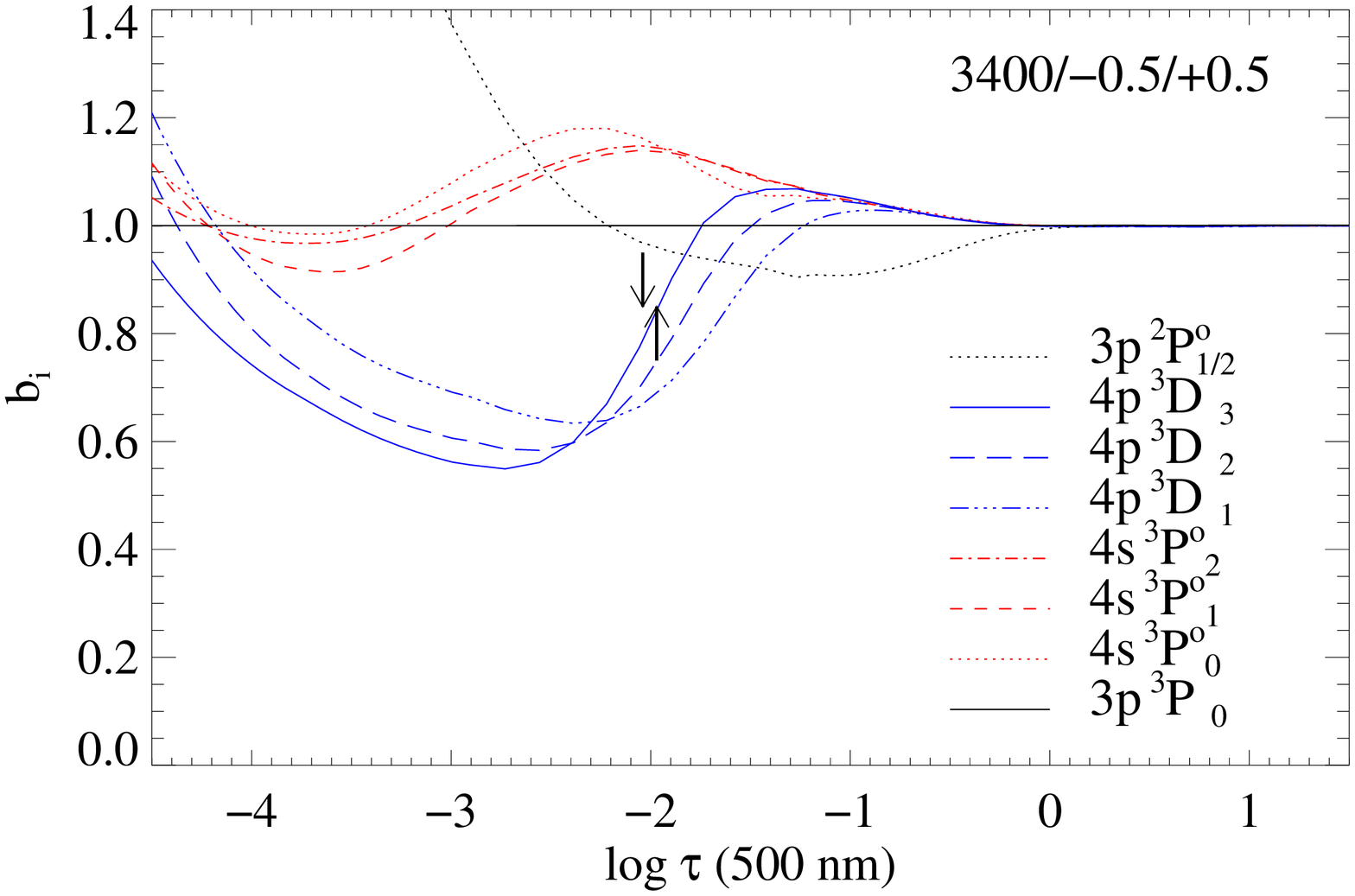}}
\caption{The NLTE departure coefficients of \sii\ for RSG models with T$_{\rm
eff} = 3400$K, log g $= -0.5$ and [Z] $= -0.5$ (top) and $0.5$ (bottom) and as a
function of optical depth. Black solid: \sii\ ground state, black dotted: \siii\
ground state. Red: lower fine structure levels of J-band IR transitions. Blue:
upper fine structure levels of IR-transitions. The LTE and NLTE line core
optical depths $\log \tau (12031$ \AA, \sii$) = 0$ are also indicated by the
upward and downward directed arrows, respectively.}
\label{dep34}
\end{center}
\end{figure*}

\begin{figure*}[ht!]
\begin{center}
\resizebox{0.6\textwidth}{!}{\includegraphics[scale=0.8]{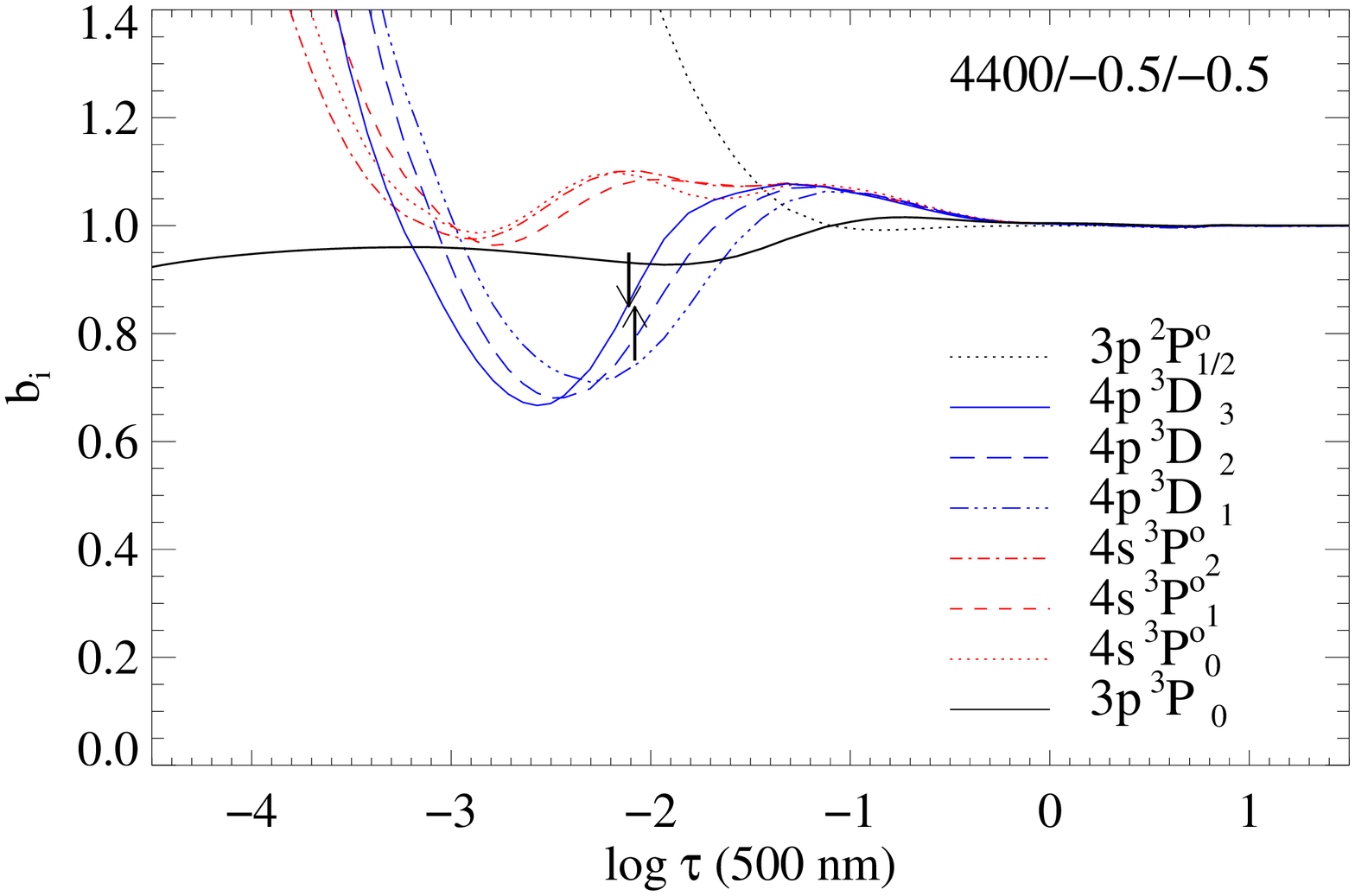}}
\resizebox{0.6\textwidth}{!}{\includegraphics[scale=0.8]{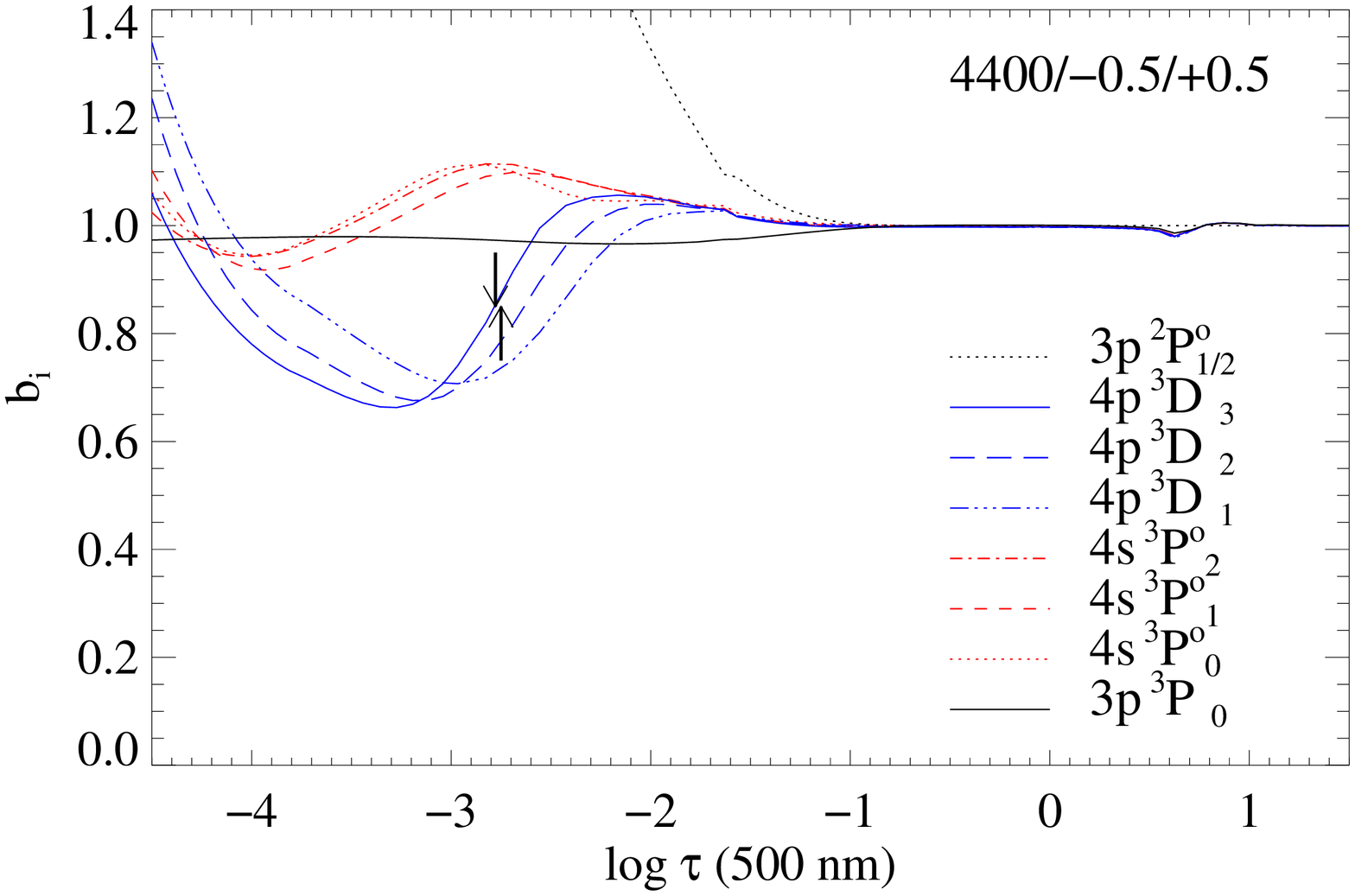}
}
\caption{Same as Fig. \ref{dep34} but for T$_{\rm eff}$ = 4400K.}
\label{dep44}
\end{center}
\end{figure*}

\section{Results}

Given the very complex atomic structure with thousands of radiative and 
collisional processes contributing to the net population or de-population
of levels a comprehensive quantitative discussion of the NLTE effects is 
complicated. For simplicity, we focus on the \sii\ levels and transitions 
related to the J-band analysis of RSG stars and on the ground states of 
\sii\ and \siii\ . Fig. \ref{dep34} and \ref{dep44} show the departure
coefficients $b_i$ of the corresponding levels defined as

\begin{equation}
b_i = n_i^{\rm NLTE}/n_i^{\rm LTE}
\end {equation}

where $n_i^{\rm NLTE}$ and $n_i^{\rm LTE}$ are NLTE and LTE atomic level
populations [cm$^{-3}$], respectively.

\begin{figure}[ht!]
\begin{center}
\resizebox{0.5\textwidth}{!}{\includegraphics[scale=1, angle=+90]
{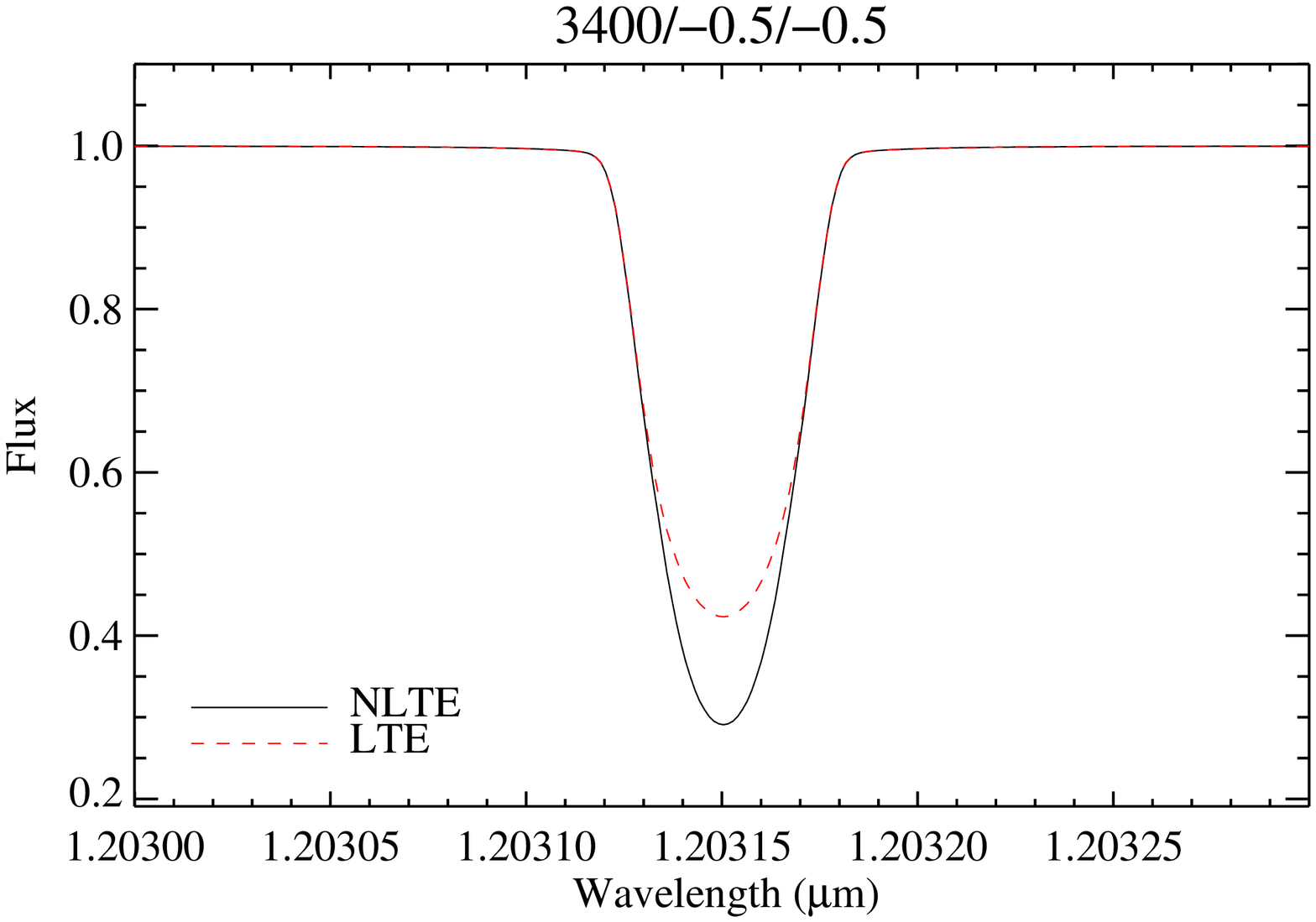}}
\resizebox{0.5\textwidth}{!}{\includegraphics[scale=1, angle=+90]
{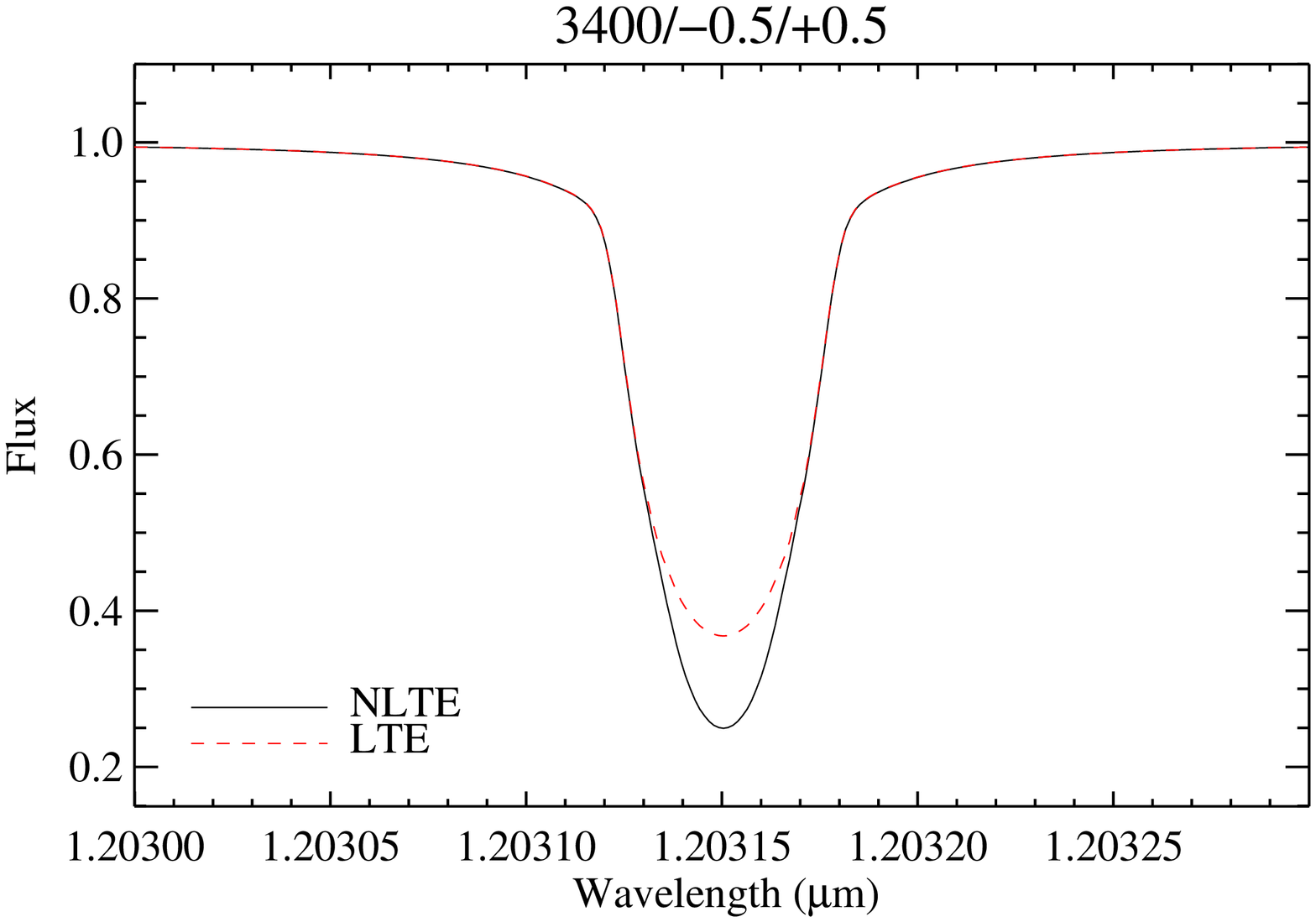}}
\caption{The NLTE (black, solid) and LTE (red, dashed) profiles of the \sii line
at $12031$ \AA\ computed for T$_{\rm eff}$ = 3400K,  log g = -0.5 and with [Z] =
-0.5 (top) and 0.5 (bottom). The  microturbulence is $\xi_{\rm t} = 2$ km/s.}
\label{prof34}
\end{center}
\end{figure}

\begin{figure}[ht!]
\begin{center}
\resizebox{0.5\textwidth}{!}{\includegraphics[scale=1, angle=+90]
{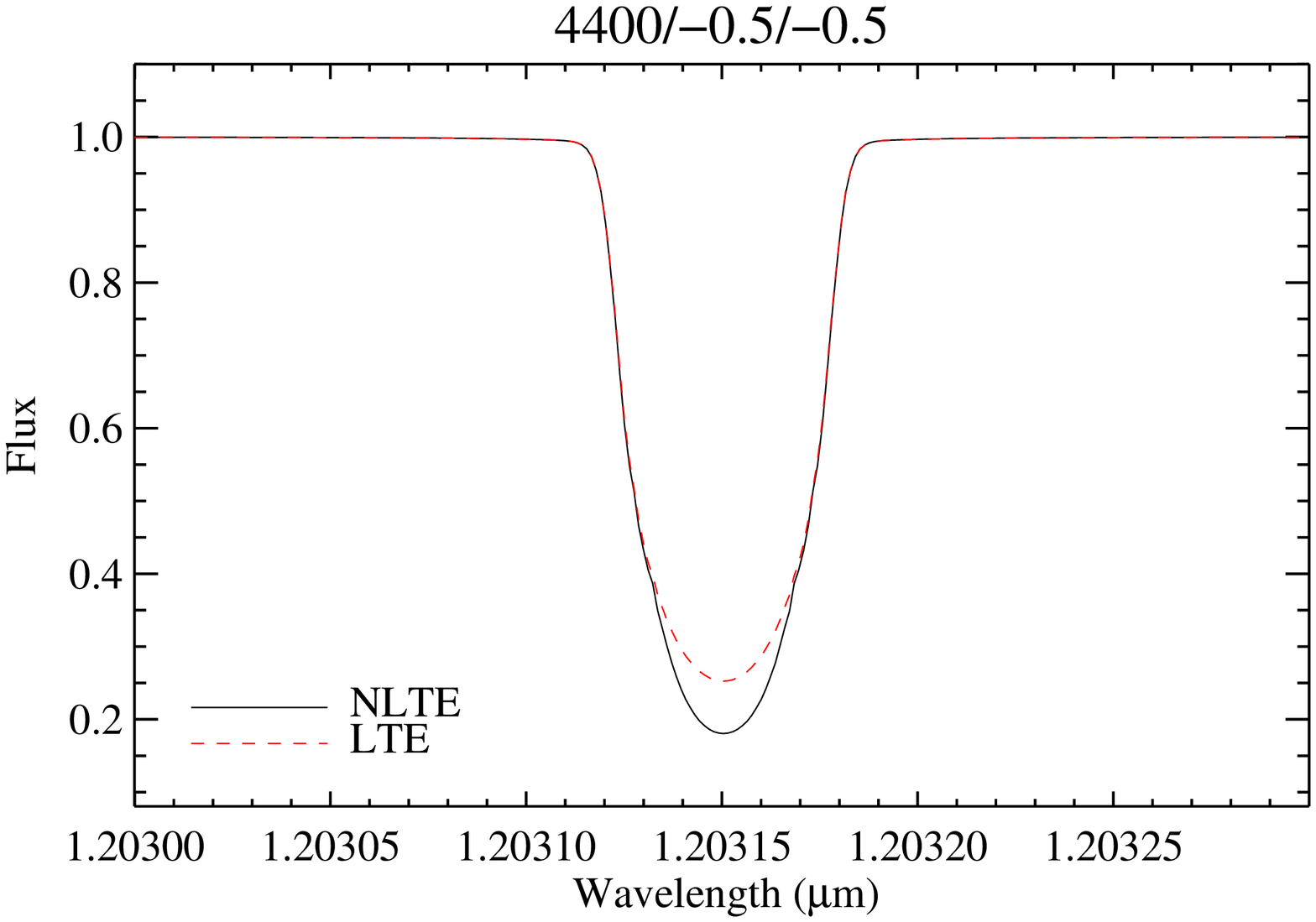}}
\resizebox{0.5\textwidth}{!}{\includegraphics[scale=1, angle=+90]
{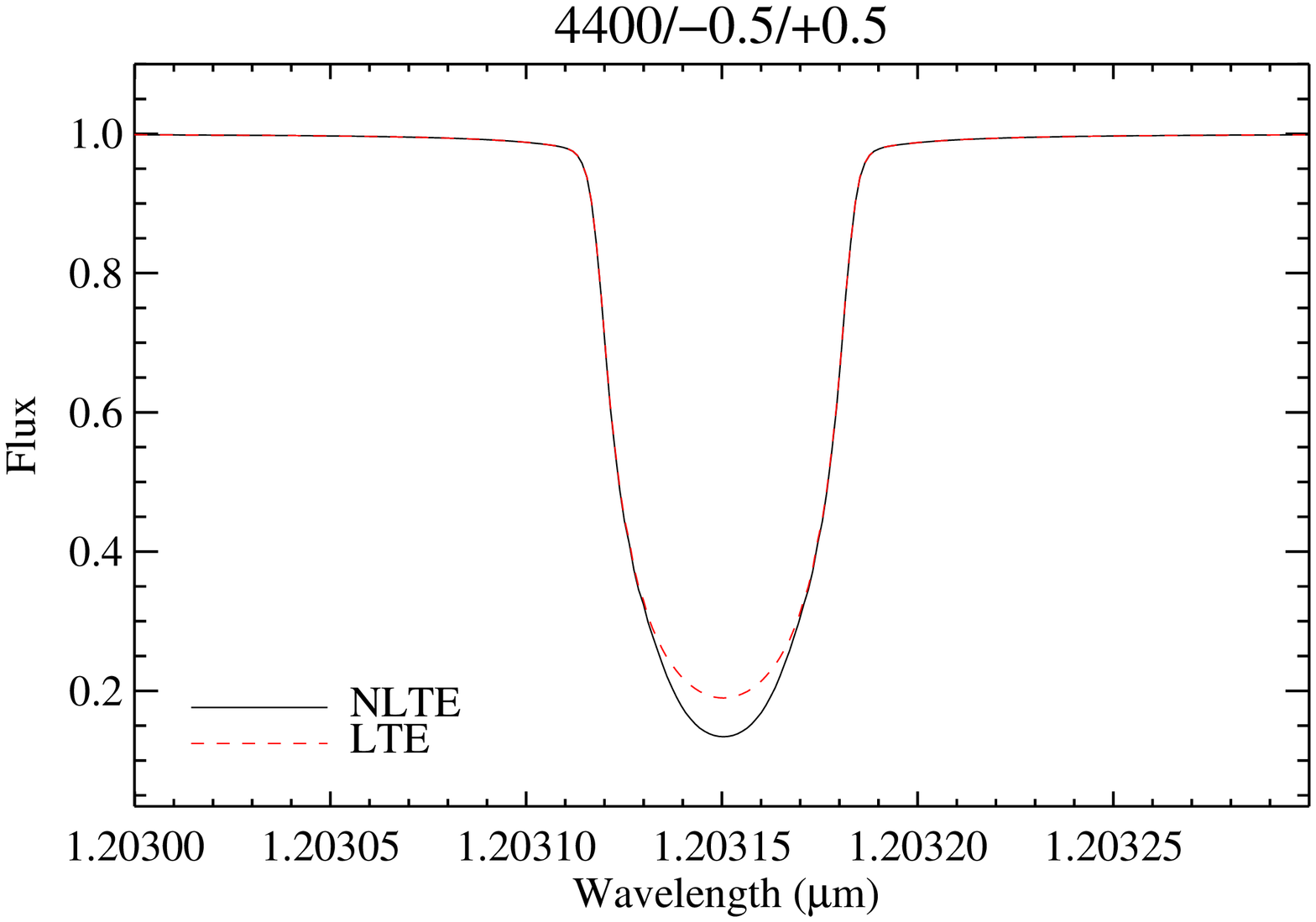}}
\caption{Same as Fig. \ref{prof34} but T$_{\rm eff}$ = 4400K.}
\label{prof44}
\end{center}
\end{figure}

\begin{figure}[ht!]
\begin{center}
\resizebox{0.45\textwidth}{!}{\includegraphics[scale=1]
{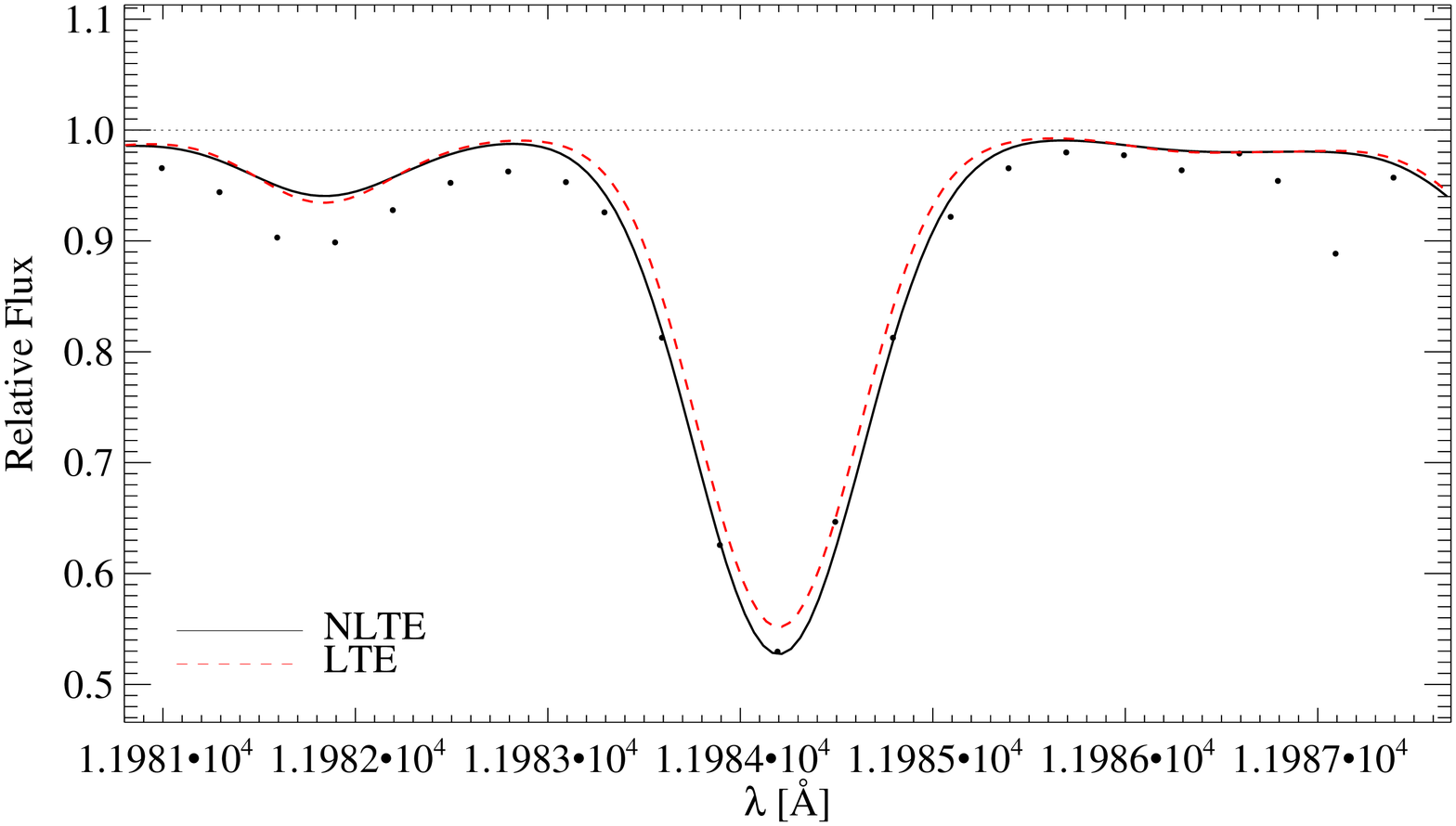}}
\resizebox{0.45\textwidth}{!}{\includegraphics[scale=1]
{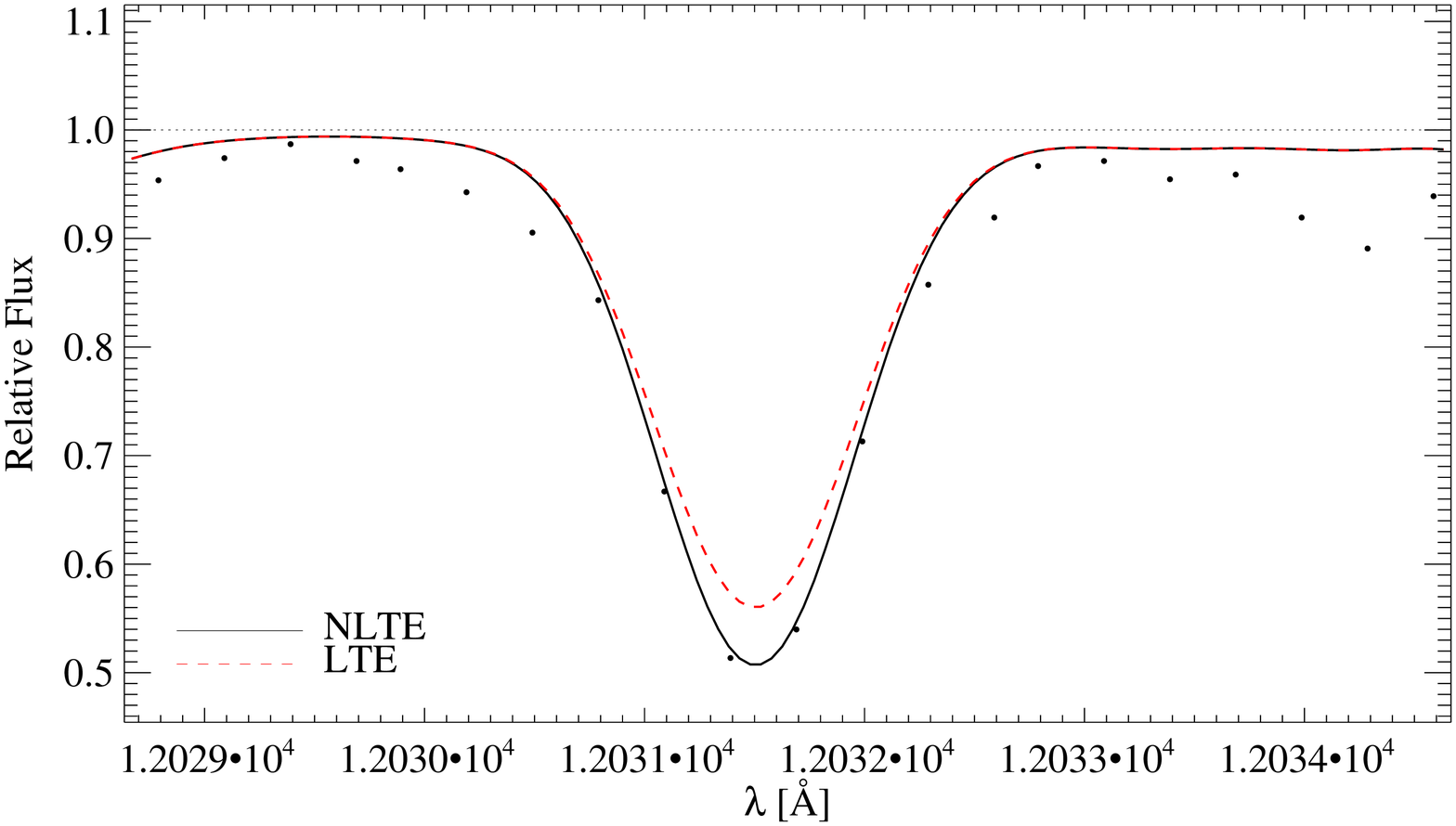}}
\resizebox{0.45\textwidth}{!}{\includegraphics[scale=1]
{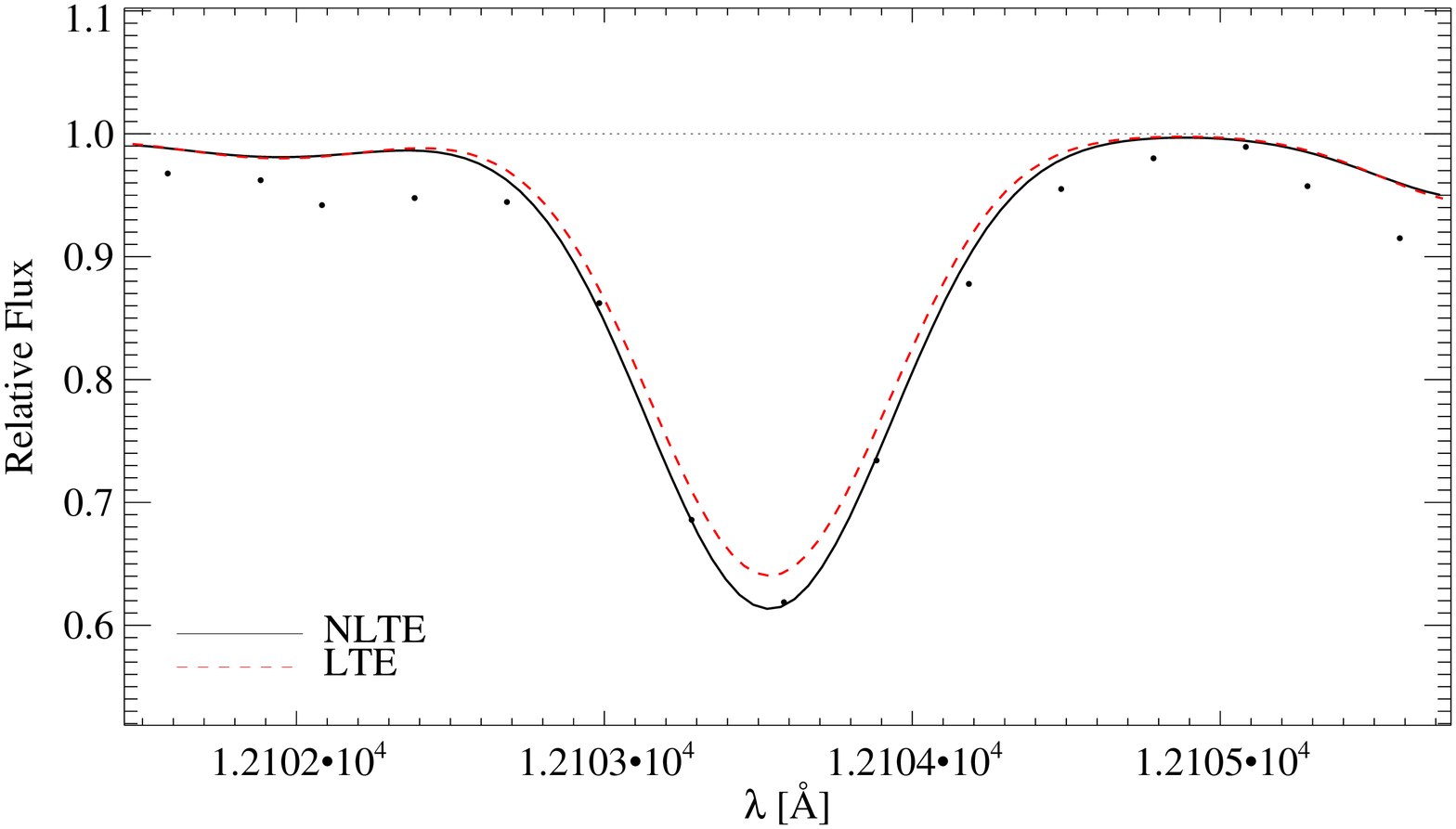}}
\caption{Subaru/IRCS high resolution observations (black dots) J-band
observations of \sii\ lines in the spectrum of the Per OB1 RSG HD 14270 (top:
11984 \AA\, middle: 12031 \AA\, bottom: 12103 \AA) compared with a LTE (red,
dashed) and a NLTE (black, solid) fit. The fit profiles have been calculated 
with $\Teff = 3800$, $\log g = 1.0$, [Z] $= 0.0$, $\xi_t = 5$ km/s.}
\label{prof_obs}
\end{center}
\end{figure}

The discussion of the NLTE effects in the lower 4s $^3$P and upper 4p $^3$D
levels of the J-band IR transitions is straightforward. The only channel 
for an allowed downward radiative decay of the lower state is towards the 
\sii\ ground state through a resonance transition. These resonance transitions
(at 2506~\AA, 2514~\AA, 2515~\AA, respectively) are optically thick
throughout the formation depths of the J-band lines. This means that downward
electron cascades from higher levels effectively stop at 4s $^3$P, which 
causes a slight overpopulation ($b_i > 1$) of these levels. At the same time
the upper levels 4p $^3$D are depopulated ($b_j < 1$)\footnote{Hereafter, $i$
and $j$ subscripts stand for the lower and upper level, respectively, and
b$_i$, b$_j$ are the corresponding departure coefficients, i.e., the
ratio of NLTE to LTE occupation numbers.} by spontaneous transitions. In
consequence, the line source function $S_{ij}$ is smaller than the local
Planck function $B_{\nu}(T_{\rm e})$ because of $S_{ij}/B_{\nu}(T_{\rm e})
\sim b_j/b_i < 1$. This weakening of the line source function (the ratio of line
emission to absorpion coefficient) together with a slight strengthening of the
line absorption coefficient through $b_i > 1$, leads to \sii\ absorption lines
which are stronger in NLTE than in LTE (Fig. \ref{prof34} and \ref{prof44}).
Fig. \ref{prof_obs} shows the NLTE and LTE line profiles in comparison
with the observations of the Per OB1 RSG HD 14270 obtained with the IRCS
spectrograph at SUBARU telescope (R $\sim 20\,000$). The fit profiles have been
calculated with $\Teff = 3800$, $\log g = 1.0$, [Z] $= 0.0$, $\xi_t = 5$ km/s.
The NLTE effects are clearly distinguishable when compared to the observed
spectrum. Therefore, one of our next steps will be to test the new
NLTE models on a large sample of high-resolution spectra of
Galactic RSG's.

From inspection of Fig. \ref{dep34} we see that for T$_{\rm eff}$ = 3400K the
departure coefficients assume values of $b_i \sim 1.15$ and $b_j\sim 0.6$ at the
depths of the formation of the line cores. These values are independent of the
silicon abundance (or metallicity [Z]). The only effect of higher abundance is
that the line cores form further out in the atmosphere, while the extreme values
of the departure coefficients remain the same. In consequence, following the 
Eddington-Barbier relationship that the emergent flux is roughly given by the
source function at optical depth $2/3$ the relation between the NLTE and LTE
emergent flux at the line center is $H^{\rm{NLTE}}_{0} \sim
\frac{b_j}{b_i}H^{\rm{LTE}}_{0} \sim 0.5 H^{\rm{LTE}}_{0}$. This is confirmed by
the actual calculations of line profiles at both metallicities (Fig.
\ref{prof34}). 

At T$_{\rm eff}$ = 4400K the NLTE effects are smaller and we obtain $b_i \sim
1.1$ and $b_j \sim 0.7,$ and  $H^{\rm{NLTE}}_{0} \sim 0.64 H^{\rm{LTE}}_{0}$
(Fig \ref{dep44} and \ref{prof44}). The reason is that at higher temperature the
rate of electron collisions populating or depopulating the higher and lower
levels increases, whereas the depopulation of the higher levels through
spontaneous emission is temperature independent.

\begin{figure*}[ht!]
\begin{center}
\includegraphics[scale=0.80]{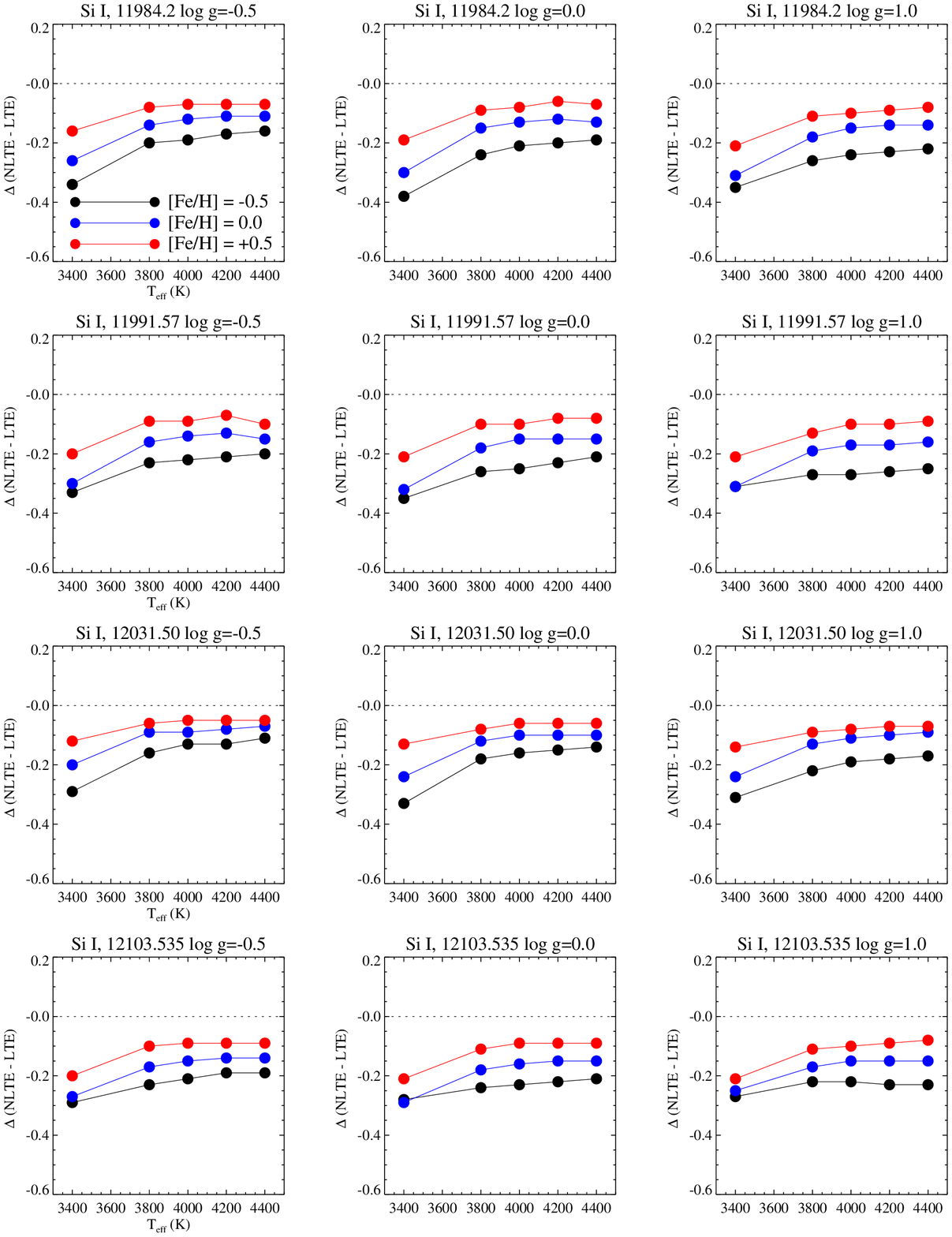}
\caption{NLTE abundance corrections as a function of effective temperature for 
microturbulence $\xi$ = 5 km/s for \sii\ 11984 \AA\ (top), 11991 \AA\ (2nd
row), 12031 \AA\ (3rd row) and 12103 \AA\ (bottom). Left column: log g =
-0.5, middle column: log g = 0.0, right column: log g = 1.0. Black solid: [Z] =
-0.5, blue solid: [Z] = 0.0, red solid: [Z] = +0.5.}
\label{nltecor5}
\end{center}
\end{figure*}

The increase of the absorption strengths of the \sii\ J-band lines in NLTE
has consequences for the determination of element abundances. The importance of
this effect can be assessed by introducing NLTE abundance corrections
$\Delta$, where:

\begin{equation}
\Delta = \log \rm{A (Si)}_{\rm NLTE} - \log \rm{A (Si)}_{\rm LTE}
\end {equation}

is the the logarithmic correction, which has to be applied to an LTE silicon
abundance determination of a specific line, $\log \rm{A}$, to obtain the correct
value corresponding to the use of NLTE line formation. We calculate these
corrections at each point of our model grid for each line by matching the NLTE
equivalent width through varying the Si abundance in the LTE calculations.
Note that from the definition of $\Delta$ a NLTE abundance correction is
negative, when for the same element abundance the NLTE line equivalent width is
larger than the LTE one, because it requires a  higher LTE abundance to fit the
NLTE equivalent width. Figs. \ref{nltecor2} and \ref{nltecor5} show the NLTE abundance corrections calculated in this way for the two values of microturbulence bracketing the values found in RSG's (Davies et al. 2010). It can be seen that 
the difference between them is small, similar to our results for the \fei\ and \tii\ J-band lines (Bergemann et al. 2012). Therefore, the data are given in Table 2 for $\xi = 2$ km/s only.

The NLTE abundance corrections are substantial with large negative values 
between $-0.4$ to $-0.1$ dex. The corrections are strongest at low silicon
abundance (or low metallicity [Z]). While as discussed above changes between LTE and NLTE central line intensity and, thus, also changes in the absolute value of
equivalent widths W$_{\lambda}$ are roughly independent of abundance (see Table 
3), relative changes of W$_{\lambda}$ are significantly larger at low abundance,
where the lines are weaker. This in turn leads to significantly larger NLTE
abundance corrections. As also discussed above NLTE effects are weaker at
higher effective temperature, which decreases the NLTE abundance corrections.

In NLTE studies, it is common to investigate the dependence of NLTE effects on inelastic collision rates with \hi. The prescriptions we adopted for the \sii\ model atom are described in Sec. \ref{sec:atom} and, to the best of our knowledge, there are presently no better alternatives. While it is unlikely that we underestimate the collision strengths \citep{barklem2011}, there is a certain degree of uncertainty with respect to their lower limit. We thus performed a series of test calculations decreasing uniformly the \hi cross-sections for all \sii\ levels by a factor of two and ten, $S_{\rm H} = 0.5, 0.1$. This scaling has a very regular effect on the NLTE corrections for different J-band \sii\ lines. In particular, for $S_{\rm H} = 0.1$, the NLTE abundance corrections (Table 2) change by $\sim -0.07$ to $-0.15$ dex at any $\Teff$ and [Z]. Clearly, more accurate estimates of the collision rates are desirable and, once available, will significantly improve the accuracy of the calculations.

\section{J-band medium resolution spectral analysis and future work}

In order to assess the influence of the NLTE effects on the J-band medium 
resolution metallicity studies, we extend the experiment already carried out in
Paper I. We calculate complete synthetic J-band spectra with MARCS model
atmospheres and LTE opacities for all spectral lines except the lines of \sii,
\fei, and \tii, for which we used our NLTE calculations. We then use
these synthetic spectra calculated for different metallicity with a fixed log g
and effective temperature (and with added Gaussian noise corresponding to S/N of
$200$) as input for the DKF10 $\chi^{2}$ analysis using MARCS model spectra
calculated completely in LTE. From the metallicities recovered in this way we
can estimate the possible systematic errors when relying on a complete LTE fit
of RSG J-band spectra.

\begin{figure*}[ht!]
\begin{center}
\includegraphics[scale=0.80]{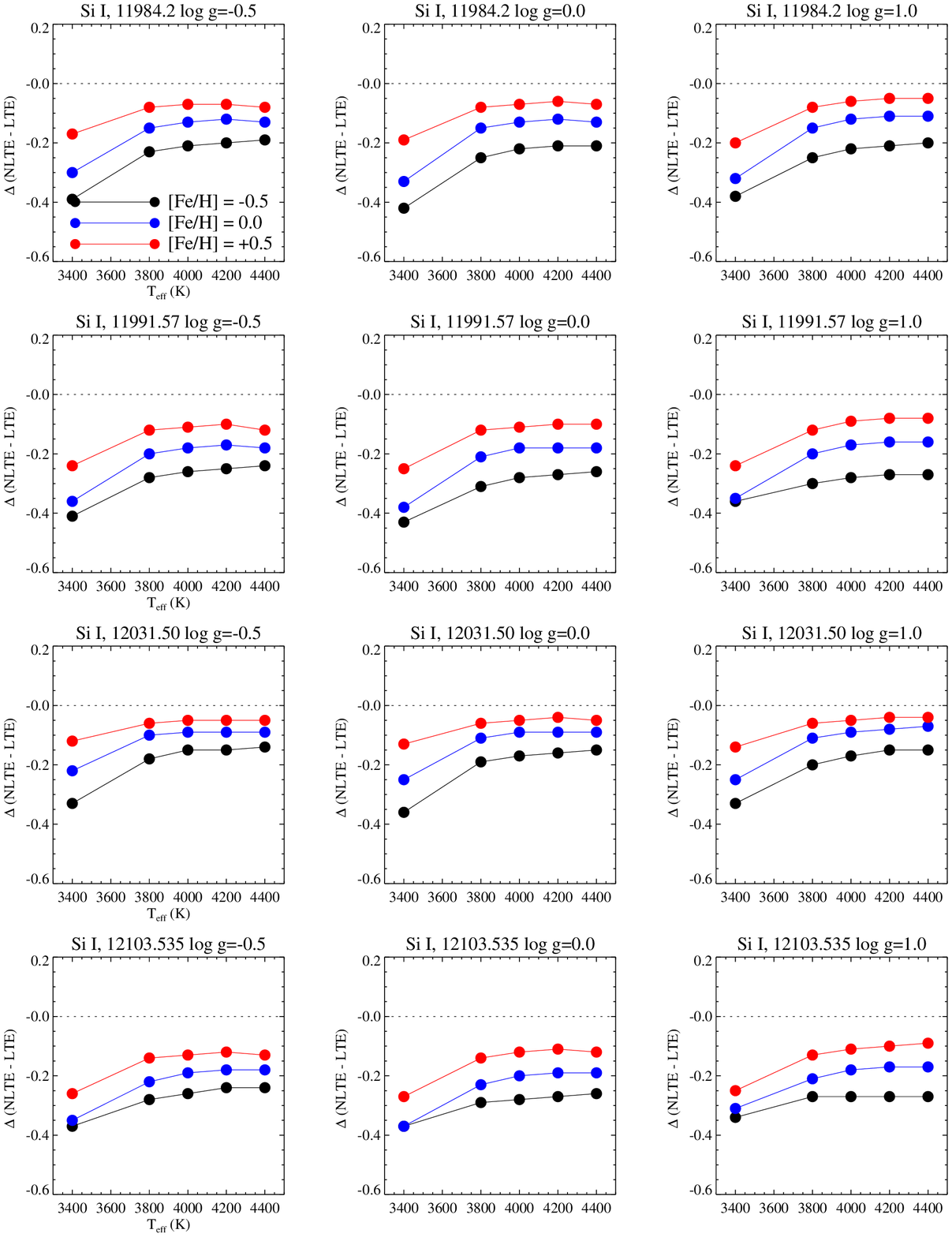}
\caption{NLTE abundance corrections as a function of effective temperature for 
microturbulence $\xi$ = 2 km/s for \sii\ 11984 \AA\ (top), 11991 \AA\ (2nd
row), 12031 \AA\ (3rd row) and 12103 \AA\ (bottom). Left column: log g =
-0.5, middle column: log g = 0.0, right column: log g = 1.0. Black solid: [Z] =
-0.5, blue solid: [Z] = 0.0, red solid: [Z] = +0.5.}
\label{nltecor2}
\end{center}
\end{figure*}

\begin{figure*}[ht!]
\begin{center}
\resizebox{0.95\textwidth}{!}{\includegraphics[scale=0.5]{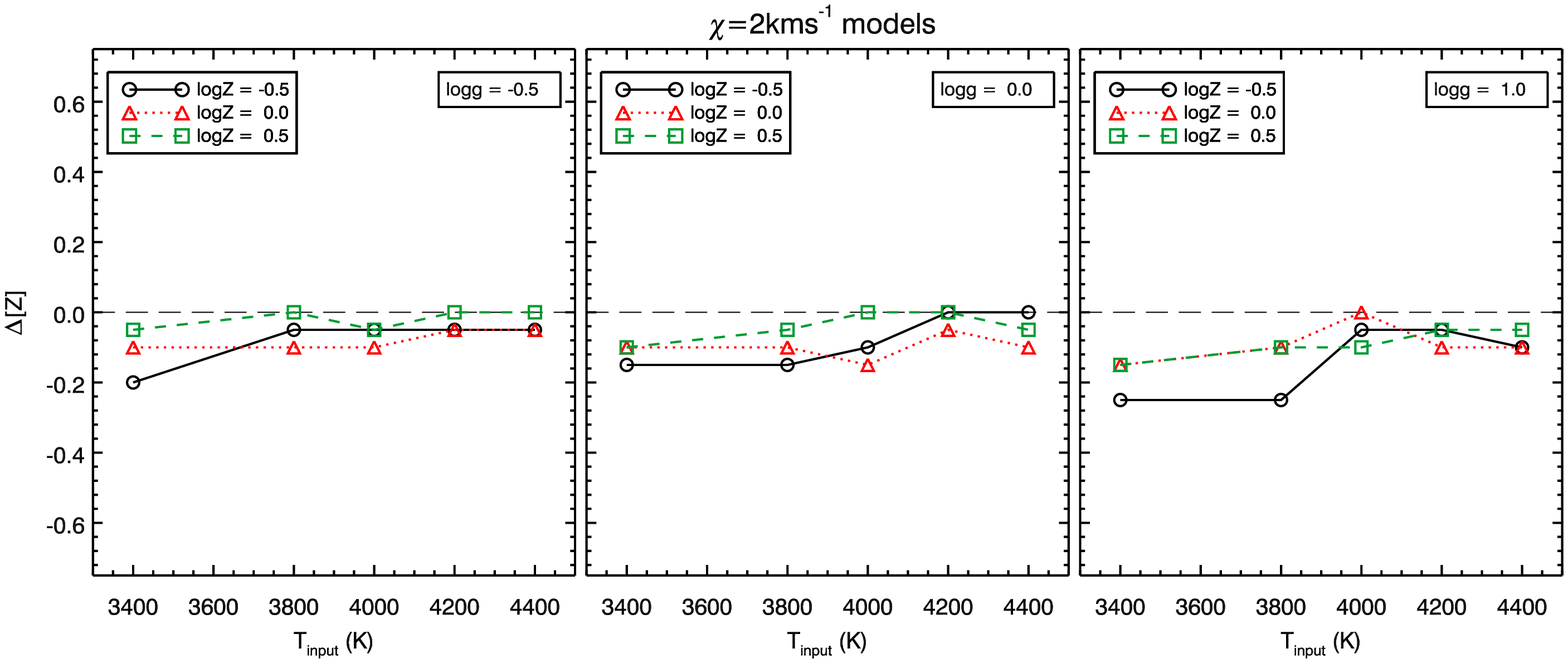}}
\caption{Influence of the \sii\ , \tii\ and \fei\ non-LTE effects on the DKF10
J-band $\chi^{2}$ metallicity determination as a function of effective
temperature. The numerical experiment is described in the text. Left: log g $=
-0.5$, middle: log g $= 0.0$, right: log g $= 1.0$. Circles: [Z] $= -0.5$,
triangles: [Z] $=0.0$, squares: [Z] $= 0.5$
}
\label{nlteparam}
\end{center}
\end{figure*}

The results of this experiment are summarized in Fig.\ref{nlteparam} and
reflect the qualitative behaviour of Fig. \ref{nltecor2} except that the total
metallicity corrections are smaller than the individual silicon abundance
corrections. This is caused by the fact that the metallicity determination with
the DKF10 method is dominated by the numerous \fei\ lines, which are relatively
well represented by the LTE approximation (see Paper I). However, at lower
effective temperature we still find corrections between $-0.15$ to $-0.25$ dex
equal or even larger than the 0.15 dex uncertainties encountered with the
DKF10 technique. In consequence, the inclusion of NLTE effects in the \sii\,
\tii\ and \fei\ lines will definitely improve the accuracy of future
extragalactic RSG J-band abundance studies.

\acknowledgments
This work was supported by the National Science Foundation under grant
AST-1108906 to RPK. Moreover, RPK acknowledges support by the
Alexander-von-Humboldt Foundation and the hospitality  of the
Max-Planck-Institute for Astrophysics in Garching and the University
Observatory Munich, where part of this work was carried out.

\clearpage

\clearpage



\begin{deluxetable}{ccccccc}
\tabletypesize{\scriptsize}
\tablecolumns{7}
\tablewidth{0pt}
\tablecaption{J-band \sii\ lines}

\tablehead{
\colhead{Elem.}     &
\colhead{$\lambda$}            &
\colhead{$\Elow$}           &
\colhead{lower}     &
\colhead{$\Eup$}     &
\colhead{upper}             &
\colhead{$log~gf$}\\
\colhead{}     &
\colhead{\AA}            &
\colhead{[eV]}           &
\colhead{level}     &
\colhead{[eV]}     &
\colhead{level}             &
\colhead{}\\[1mm]
\colhead{(1)}	&
\colhead{(2)}	&
\colhead{(3)}	&
\colhead{(4)}	&
\colhead{(5)}	&
\colhead{(6)}	&
\colhead{(7)}}	
\startdata
\\[-1mm]
\sii & 11991.57 & 4.92 & 4s $^3$P$_0^{\mathrm o}$ & 5.95 & 4p $^3$D$_1$ & -0.109 \\
     & 11984.20 & 4.93 & 4s $^3$P$_1^{\mathrm o}$ & 5.96 & 4p $^3$D$_2$ &  0.239 \\
     & 12103.54 & 4.93 & 4s $^3$P$_1^{\mathrm o}$ & 5.95 & 4p $^3$D$_1$ & -0.351 \\
     & 12031.50 & 4.95 & 4s $^3$P$_2^{\mathrm o}$ & 5.98 & 4p $^3$D$_3$ &  0.477 \\

\enddata
\end{deluxetable}

\clearpage

%
%
\begin{deluxetable}{ccccccc}
\tabletypesize{\scriptsize}
\tablecolumns{9}
\tablewidth{0pt}
\tablecaption{Non-LTE abundance corrections for the \sii\ lines ($\xi$ = 2 kms$^{-1}$)}

\tablehead{
\colhead{T$_{\rm eff}$}     &
\colhead{$log~g$}            &
\colhead{[Z]}           &
\colhead{$\Delta_{\rm Si I}$}     &
\colhead{$\Delta_{\rm Si I}$}             &
\colhead{$\Delta_{\rm Si I}$}     &
\colhead{$\Delta_{\rm Si I}$}\\
\colhead{}     &
\colhead{}            &
\colhead{}           &
\colhead{$11984.20$}     &
\colhead{$11991.57$}     &
\colhead{$12031.50$}             &
\colhead{$12103.54$} \\[1mm]
\colhead{(1)}	&
\colhead{(2)}	&
\colhead{(3)}	&
\colhead{(4)}	&
\colhead{(5)}	&
\colhead{(6)}	&
\colhead{(7)}    }	
\startdata
\\[-1mm]
 4400.  & -0.50  &  0.50 &  -0.08 &  -0.12 &  -0.05 &  -0.13 \\
 4400.  & -0.50  &  0.00 &  -0.13 &  -0.18 &  -0.09 &  -0.18 \\
 4400.  & -0.50  & -0.50 &  -0.19 &  -0.24 &  -0.14 &  -0.24 \\
 4400.  &  0.00  &  0.50 &  -0.07 &  -0.10 &  -0.05 &  -0.12 \\
 4400.  &  0.00  &  0.00 &  -0.13 &  -0.18 &  -0.09 &  -0.19 \\
 4400.  &  0.00  & -0.50 &  -0.21 &  -0.26 &  -0.15 &  -0.26 \\
 4400.  &  1.00  &  0.50 &  -0.05 &  -0.08 &  -0.04 &  -0.09 \\
 4400.  &  1.00  &  0.00 &  -0.11 &  -0.16 &  -0.07 &  -0.17 \\
 4400.  &  1.00  & -0.50 &  -0.20 &  -0.27 &  -0.15 &  -0.27 \\
 & & & & & & \\
 4200.  & -0.50  &  0.50 &  -0.07 &  -0.10 &  -0.05 &  -0.12 \\
 4200.  & -0.50  &  0.00 &  -0.12 &  -0.17 &  -0.09 &  -0.18 \\
 4200.  & -0.50  & -0.50 &  -0.20 &  -0.25 &  -0.15 &  -0.24 \\
 4200.  &  0.00  &  0.50 &  -0.06 &  -0.10 &  -0.04 &  -0.11 \\
 4200.  &  0.00  &  0.00 &  -0.12 &  -0.18 &  -0.09 &  -0.19 \\
 4200.  &  0.00  & -0.50 &  -0.21 &  -0.27 &  -0.16 &  -0.27 \\
 4200.  &  1.00  &  0.50 &  -0.05 &  -0.08 &  -0.04 &  -0.10 \\
 4200.  &  1.00  &  0.00 &  -0.11 &  -0.16 &  -0.08 &  -0.17 \\
 4200.  &  1.00  & -0.50 &  -0.21 &  -0.27 &  -0.15 &  -0.27 \\
 & & & & & & \\
 4000.  & -0.50  &  0.50 &  -0.07 &  -0.11 &  -0.05 &  -0.13 \\
 4000.  & -0.50  &  0.00 &  -0.13 &  -0.18 &  -0.09 &  -0.19 \\
 4000.  & -0.50  & -0.50 &  -0.21 &  -0.26 &  -0.15 &  -0.26 \\
 4000.  &  0.00  &  0.50 &  -0.07 &  -0.11 &  -0.05 &  -0.12 \\
 4000.  &  0.00  &  0.00 &  -0.13 &  -0.18 &  -0.09 &  -0.20 \\
 4000.  &  0.00  & -0.50 &  -0.22 &  -0.28 &  -0.17 &  -0.28 \\
 4000.  &  1.00  &  0.50 &  -0.06 &  -0.09 &  -0.05 &  -0.11 \\
 4000.  &  1.00  &  0.00 &  -0.12 &  -0.17 &  -0.09 &  -0.18 \\
 4000.  &  1.00  & -0.50 &  -0.22 &  -0.28 &  -0.17 &  -0.27 \\
 & & & & & & \\
 3800.  & -0.50  &  0.50 &  -0.08 &  -0.12 &  -0.06 &  -0.14 \\
 3800.  & -0.50  &  0.00 &  -0.15 &  -0.20 &  -0.10 &  -0.22 \\
 3800.  & -0.50  & -0.50 &  -0.23 &  -0.28 &  -0.18 &  -0.28 \\
 3800.  &  1.00  &  0.50 &  -0.08 &  -0.12 &  -0.06 &  -0.13 \\
 3800.  &  1.00  &  0.00 &  -0.15 &  -0.20 &  -0.11 &  -0.21 \\
 3800.  &  1.00  & -0.50 &  -0.25 &  -0.30 &  -0.20 &  -0.27 \\
 3800.  &  0.00  &  0.50 &  -0.08 &  -0.12 &  -0.06 &  -0.14 \\
 3800.  &  0.00  &  0.00 &  -0.15 &  -0.21 &  -0.11 &  -0.23 \\
 3800.  &  0.00  & -0.50 &  -0.25 &  -0.31 &  -0.19 &  -0.29 \\
 & & & & & & \\
 3400.  & -0.50  &  0.50 &  -0.17 &  -0.24 &  -0.12 &  -0.26 \\
 3400.  & -0.50  &  0.00 &  -0.30 &  -0.36 &  -0.22 &  -0.35 \\
 3400.  & -0.50  & -0.50 &  -0.39 &  -0.41 &  -0.33 &  -0.37 \\
 3400.  &  0.00  &  0.50 &  -0.19 &  -0.25 &  -0.13 &  -0.27 \\
 3400.  &  0.00  &  0.00 &  -0.33 &  -0.38 &  -0.25 &  -0.37 \\
 3400.  &  0.00  & -0.50 &  -0.42 &  -0.43 &  -0.36 &  -0.37 \\
 3400.  &  1.00  &  0.50 &  -0.20 &  -0.24 &  -0.14 &  -0.25 \\
 3400.  &  1.00  &  0.00 &  -0.32 &  -0.35 &  -0.25 &  -0.31 \\
 3400.  &  1.00  & -0.50 &  -0.38 &  -0.36 &  -0.33 &  -0.34 \\
\enddata
\end{deluxetable}

\clearpage

%
%

\begin{deluxetable}{ccccccccccc}
\tabletypesize{\scriptsize}
\tablecolumns{11}
\tablewidth{0pt}
\tablecaption{Equivalent widths \tablenotemark{a} of the \sii\ lines ($\xi$ = 2 kms$^{-1}$)}

\tablehead{
\colhead{T$_{\rm eff}$}     &
\colhead{$log~g$}          &
\colhead{[Z]}              &
\colhead{$W_{\lambda,\rm Si I}$}     &
\colhead{$W_{\lambda,\rm Si I}$}     &
\colhead{$W_{\lambda,\rm Si I}$}     &
\colhead{$W_{\lambda,\rm Si I}$}     &
\colhead{$W_{\lambda,\rm Si I}$}     &
\colhead{$W_{\lambda,\rm Si I}$}     &
\colhead{$W_{\lambda,\rm Si I}$}     &
\colhead{$W_{\lambda,\rm Si I}$}     \\
\colhead{}            &
\colhead{}            &
\colhead{}            &
\colhead{$11984$}             &
\colhead{$11984$}             &
\colhead{$11991$}     &
\colhead{$11991$}     &
\colhead{$12031$}     &
\colhead{$12031$}     &
\colhead{$12103$}     &
\colhead{$12103$}\\[1mm]
\colhead{}            &
\colhead{}            &
\colhead{}            &
\colhead{$LTE$}       &
\colhead{$NLTE$}      &
\colhead{$LTE$}       &
\colhead{$NLTE$}      &
\colhead{$LTE$}       &
\colhead{$NLTE$}      &
\colhead{$LTE$}       &
\colhead{$NLTE$}\\[1mm]
\colhead{(1)}	&
\colhead{(2)}	&
\colhead{(3)}	&
\colhead{(4)}	&
\colhead{(5)}	&
\colhead{(6)}	&
\colhead{(7)}   &
\colhead{(8)}	&
\colhead{(9)}	&
\colhead{(10)}  &
\colhead{(11)}}
\startdata
\\[-1mm]
4400 & -0.50 &  0.00 & 381.3 & 402.9 & 350.5 & 373.6 & 403.6 & 422.6 & 330.6 & 351.7 \\
4400 & -0.50 &  0.50 & 426.2 & 444.8 & 392.5 & 412.3 & 452.3 & 468.7 & 372.0 & 390.0 \\
4400 & -0.50 & -0.50 & 340.5 & 365.3 & 309.7 & 336.4 & 362.2 & 383.6 & 289.4 & 314.0 \\
4400 &  0.00 &  0.00 & 369.4 & 393.8 & 337.4 & 363.3 & 393.3 & 414.8 & 317.3 & 340.8 \\
4400 &  0.00 &  0.50 & 422.4 & 443.2 & 384.3 & 406.3 & 453.5 & 471.9 & 362.2 & 382.1 \\
4400 &  0.00 & -0.50 & 327.0 & 355.4 & 296.1 & 326.3 & 349.2 & 373.9 & 276.2 & 303.6 \\
4400 &  1.00 &  0.00 & 348.9 & 378.3 & 311.6 & 342.3 & 379.0 & 405.0 & 289.9 & 317.2 \\
4400 &  1.00 &  0.50 & 427.7 & 452.8 & 371.9 & 398.1 & 478.5 & 501.0 & 343.0 & 366.5 \\
4400 &  1.00 & -0.50 & 299.5 & 334.0 & 267.6 & 303.2 & 323.6 & 354.0 & 247.9 & 278.9 \\
& & & & & & & & & & \\
4200 & -0.50 &  0.00 & 375.6 & 397.9 & 345.1 & 368.9 & 397.6 & 417.2 & 325.6 & 347.4 \\
4200 & -0.50 &  0.50 & 422.6 & 442.0 & 388.3 & 408.9 & 449.5 & 466.7 & 367.7 & 386.5 \\
4200 & -0.50 & -0.50 & 335.9 & 361.7 & 306.0 & 333.5 & 356.9 & 379.4 & 286.3 & 311.5 \\
4200 &  0.00 &  0.00 & 360.0 & 385.1 & 328.3 & 354.9 & 383.6 & 405.6 & 308.5 & 332.7 \\
4200 &  0.00 &  0.50 & 415.4 & 437.3 & 376.2 & 399.2 & 448.0 & 467.4 & 353.9 & 374.8 \\
4200 &  0.00 & -0.50 & 317.8 & 347.0 & 287.7 & 318.7 & 339.1 & 364.6 & 268.3 & 296.3 \\
4200 &  1.00 &  0.00 & 332.3 & 362.8 & 295.8 & 327.4 & 361.7 & 388.7 & 274.6 & 302.4 \\
4200 &  1.00 &  0.50 & 412.6 & 439.8 & 356.2 & 384.3 & 464.4 & 488.7 & 327.1 & 352.2 \\
4200 &  1.00 & -0.50 & 283.9 & 318.9 & 253.3 & 288.9 & 306.7 & 337.8 & 234.2 & 264.8 \\
& & & & & & & & & & \\
4000 & -0.50 &  0.00 & 363.6 & 387.9 & 330.2 & 355.8 & 389.3 & 410.7 & 309.7 & 333.1 \\
4000 & -0.50 &  0.50 & 423.3 & 445.8 & 377.2 & 400.7 & 463.6 & 483.6 & 351.9 & 373.4 \\
4000 & -0.50 & -0.50 & 321.7 & 348.9 & 291.4 & 320.2 & 343.2 & 367.0 & 271.8 & 298.1 \\
4000 &  0.00 &  0.00 & 349.1 & 376.5 & 312.4 & 341.0 & 378.6 & 402.8 & 290.7 & 316.7 \\
4000 &  0.00 &  0.50 & 425.5 & 451.1 & 368.2 & 394.8 & 477.9 & 500.8 & 338.4 & 362.6 \\
4000 &  0.00 & -0.50 & 301.9 & 332.6 & 271.1 & 303.2 & 324.5 & 351.3 & 251.6 & 280.4 \\
4000 &  1.00 &  0.00 & 329.4 & 362.9 & 282.5 & 316.5 & 370.3 & 400.3 & 257.1 & 286.5 \\
4000 &  1.00 &  0.50 & 450.1 & 482.2 & 362.8 & 395.5 & 530.4 & 559.3 & 320.0 & 348.8 \\
4000 &  1.00 & -0.50 & 261.9 & 297.8 & 232.5 & 268.3 & 283.4 & 315.8 & 213.9 & 243.8 \\
& & & & & & & & & & \\
3800 & -0.50 &  0.00 & 343.2 & 371.2 & 306.6 & 335.5 & 372.7 & 397.4 & 285.0 & 311.4 \\
3800 & -0.50 &  0.50 & 415.4 & 442.5 & 357.7 & 385.6 & 468.0 & 492.0 & 327.6 & 353.1 \\
3800 & -0.50 & -0.50 & 299.0 & 328.6 & 268.5 & 299.2 & 321.0 & 347.0 & 248.9 & 276.7 \\
3800 &  0.00 &  0.00 & 329.2 & 360.8 & 288.2 & 320.6 & 363.7 & 391.7 & 265.1 & 294.2 \\
3800 &  0.00 &  0.50 & 420.6 & 451.2 & 349.9 & 381.4 & 485.2 & 512.3 & 314.1 & 342.6 \\
3800 &  0.00 & -0.50 & 278.8 & 311.8 & 247.5 & 281.1 & 302.1 & 331.4 & 227.9 & 257.5 \\
3800 &  1.00 &  0.00 & 293.3 & 330.8 & 250.1 & 287.0 & 330.5 & 364.4 & 226.5 & 257.4 \\
3800 &  1.00 &  0.50 & 402.1 & 438.6 & 323.4 & 360.0 & 474.3 & 506.9 & 284.4 & 316.3 \\
3800 &  1.00 & -0.50 & 238.9 & 276.1 & 207.7 & 243.3 & 262.6 & 297.3 & 188.4 & 216.8 \\
& & & & & & & & & & \\
3400 & -0.50 &  0.00 & 260.8 & 298.9 & 228.6 & 265.9 & 286.3 & 319.1 & 209.5 & 243.1 \\
3400 & -0.50 &  0.50 & 324.6 & 360.8 & 275.8 & 311.9 & 367.2 & 398.2 & 249.6 & 283.3 \\
3400 & -0.50 & -0.50 & 222.8 & 260.7 & 195.6 & 231.8 & 242.2 & 276.5 & 178.1 & 209.4 \\
3400 &  0.00 &  0.00 & 240.6 & 281.6 & 206.7 & 245.8 & 267.9 & 303.8 & 187.0 & 221.2 \\
3400 &  0.00 &  0.50 & 311.2 & 351.0 & 257.8 & 296.8 & 358.5 & 392.4 & 229.6 & 265.1 \\
3400 &  0.00 & -0.50 & 200.1 & 239.5 & 172.9 & 209.4 & 219.9 & 256.3 & 155.6 & 185.6 \\
3400 &  1.00 &  0.00 & 204.8 & 245.3 & 167.4 & 203.2 & 236.2 & 273.4 & 146.6 & 174.9 \\
3400 &  1.00 &  0.50 & 285.1 & 328.7 & 224.1 & 264.1 & 340.0 & 378.3 & 192.2 & 225.9 \\
3400 &  1.00 & -0.50 & 160.5 & 194.2 & 133.2 & 161.9 & 181.3 & 214.4 & 116.5 & 137.8 \\
\\
\enddata
\tablenotetext{a}{equivalent widths $\EW$ are given in \mA}
\end{deluxetable}

\clearpage



\end{document}